\documentclass[12pt]{iopart}

\usepackage{iopams} 

\usepackage{xcolor}

\usepackage{graphicx}

\begin{document}

\title[Finite-Time Analysis of Crises in a Chaotically Forced Ocean Model]{Finite-Time Analysis of Crises in a Chaotically Forced Ocean Model}

\author{Andrew R Axelsen, Courtney R Quinn, Andrew P Bassom}

\address{School of Natural Sciences, University of Tasmania, Churchill Avenue, Sandy Bay, 7001, Tasmania, Australia}
\ead{courtney.quinn@utas.edu.au}
\vspace{10pt}
\begin{indented}
\item[]April 2024
\end{indented}

\begin{abstract}
We consider a coupling of the Stommel box model and the Lorenz model, with the goal of investigating the so-called ``crises" that are known to occur given sufficient forcing. In this context, a crisis is characterized as the destruction of a chaotic attractor under a critical forcing strength. We document the variety of chaotic attractors and crises possible in our model, focusing on the parameter region where the Lorenz model is always chaotic and where bistability exists in the Stommel box model. The chaotic saddle collisions that occur in a boundary crisis are visualized, with the chaotic saddle computed using the Saddle-Straddle Algorithm. We identify a novel sub-type of boundary crisis, namely a vanishing basin crisis.  For forcing strength beyond the crisis, we demonstrate the possibility of a merging between the persisting chaotic attractor and either a chaotic transient or a ghost attractor depending on the type of boundary crisis. An investigation of the finite-time Lyapunov exponents around crisis levels of forcing reveals a convergence between two near-neutral exponents, particularly at points of a trajectory most sensitive to divergence. This points to loss of hyperbolicity associated with crisis occurrence. Finally, we generalize our findings by coupling the Stommel box model to other strange attractors and thereby show that the behaviours are quite generic and robust.
\end{abstract}

%
\vspace{2pc}


%
\maketitle
%
%

\section{Introduction}

\label{sec:intro}
The Stommel Box Model \cite{stommel1961thermohaline} and the Lorenz model \cite{lorenz1963deterministic} (hereafter referred to as SBM and L63) are two fundamental conceptual climate models that represent surface fluxes in the North Atlantic and atmospheric convection, respectively. The SBM has proved to be a useful idealisation of the mathematics that underpins the principal mechanisms at play within the North Atlantic Ocean, while the L63 model has become recognized as a sytem which is ubiquitous in the world of chaotic dynamics. More recently the L63 model has been used to add chaotic forcing to the SBM dynamics \cite{ashwin2021physical} and this idea forms the cornerstone of the present study. Before we consider the interplay of the two models, we recall some of their individual key properties.

The SBM was suggested as an approach to assessing the surface flux variance in the North Atlantic Ocean. There are two major flux variances to consider. One of these, the thermal water flux, transports warm, saline water from the equatorial to polar regions. The other is the freshwater flux, in which fresh water is transported from the polar towards equatorial regions. The main aim of the model is to assess the behaviour when the two opposing surface fluxes meet through a two-box approach as explained by Dijkstra and Ghil \cite{dijkstra2005low}. A variety of refinements to the standard SBM have been proposed  (e.g. \cite{stommel1961thermohaline}, \cite{lohmann1999dynamics}), but for the purposes of our work, we adopt the form used in \cite{ashwin2021physical} and \cite{dijkstra2005low}. Consequently, we suppose that the quantities $T$ and $S$ denote non-dimensional variables related to the temperature and salinity differences between the equatorial and polar regions. They are coupled according to the system
\begin{equation}
    \begin{array}{lcl}
        \dot{T}(t) & = & \xi - T(1 + |T - S|), \\
        \dot{S}(t) & = & \eta - S(\zeta + |T - S|),
    \end{array}
    \label{eqn:SBM}
\end{equation}
in which the variable $t$ is a suitably dimensionless time and $\dot{[]}$ refers to the time derivative. Three key constants arise: $\xi$ and $\eta$ represent the constant rates of forcing of $T$ and $S$, while $\zeta$ is a dimensionless ratio of the thermal and salinity timescales. We also remark that when $T$ $>$ $S$, we say that the model is in a thermal-driven (TH) state, while $T$ $<$ $S$ indicates a saline-driven (SA) state \cite{dijkstra2005low}. This model possesses two principal modes of stability: one occurs in the form of monostability (one stable equilibrium) while the alternative is bistability (two stable equilibria and a saddle which determine the basins of attraction). In the bistable case, one equilibrium is in the TH state while the other is in the SA state. Furthermore, it is possible that some very special cases of stability may occur; for instance, for particular combinations of parameters there can be two stable equilibria with one of these on the $T$ $=$ $S$ line whose basin of attraction consists of a single point - itself.

The L63 system was proposed in \cite{lorenz1963deterministic} as a simple model of atmospheric convection that relates various physical properties of a two-dimensional layer that is warmed from below and cooled from above. With certain choices of parameters, the model can create the chaotic attractor that is arguably the most famous example of its type. The model is given by
\begin{equation}
    \begin{array}{lcl}
        \dot{x}(t) & = & \mu (y - x), \\
        \dot{y}(t) & = & x(\rho - z) - y, \\
        \dot{z}(t) & = & xy - \beta z,
    \end{array}
    \label{eqn:L63Model}
\end{equation}
where $x$, $y$, $z$ $\in$ $\mathbb{R}$ are variables, while the positive parameters $\mu$, $\rho$ and $\beta$ are fixed. For the purposes of this study, we specify $(\mu, \rho, \beta) = (10, 28, \frac{8}{3})$; a combination for which a well-known chaotic attractor exists \cite{lorenz1963deterministic}.

In a study by Ashwin and Newman \cite{ashwin2021physical}, which focused on measures for pullback attractors and tipping point probabilities, the authors combined these two models into what we shall subsequently refer to as the Lorenz-Forced Stommel Box Model (LFSBM). In this, the SBM (\ref{eqn:SBM}) is forced by the L63 model (\ref{eqn:L63Model}), which is used solely as a device for introducing chaotic variability. This is done by adding a forcing term to the SBM equations which is proportional to $x$-component of the Lorenz model. Following \cite{ashwin2021physical}, we are left with the combined system
\begin{equation}
    \begin{array}{lcl}
        \dot{x}(t) & = & \mu (y - x), \\
        \dot{y}(t) & = & x(\rho - z) - y, \\
        \dot{z}(t) & = & xy - \beta z, \\
        \dot{T}(t) & = & \xi + ax - T(1 + |T - S|), \\
        \dot{S}(t) & = & \eta + ax - S(\zeta + |T - S|).
    \end{array}
    \label{eqn:LFSBM}
\end{equation}
This system is an example of a technique often used to analyze non-autonomous systems whereby forced system is converted into a fully autonomous version by expressing the forcing as a stand alone ODE system which is then augmented by the appropriate number of equations \cite{ben2021useful}. For applications, it is worth pointing out that the L63 and SBM operate independently on different timescales, so that there are distinct separated timescales at play within the LFSBM. We choose equal timescales for the purpose of this study.

We focus on the SBM parameter regime that exhibits bistability. With weak forcing ($a>0$ but small), we expect the stable equilibiria of the SBM model to become chaotic attractors in the LFSBM extension and the unstable equilibrium to become a chaotic saddle. This implies the bistability is preserved between the two attractors. As the forcing strength is increased, a critical threshold is reached at which the system undergoes a so-called ``crisis" where the bistability in the system is reduced to monostability. Our focus in this work is to probe the properties of such crisis events. 

The rest of the paper is organized as follows. We commence in Section \ref{sec:behaviour} with an analysis of the crises that arise in the LFSBM and monitor the evolution of surviving chaotic attractors as the forcing increases further. Section \ref{sec:results} is devoted to an investigation of the behaviour of solutions both prior and subsequent to the occurrence of a crisis; we tackle this using a finite-time analysis. Section \ref{sec:generality} considers how our findings might have generality beyond our specific model. Section \ref{sec:conclusion} provides a few final remarks while more technical aspects of the numerical methods are relegated to the Appendix.

\section{Behaviour of the chaotically forced model\label{sec:behaviour}}

\begin{figure}
    \centering
    \includegraphics[width=0.4955\textwidth]{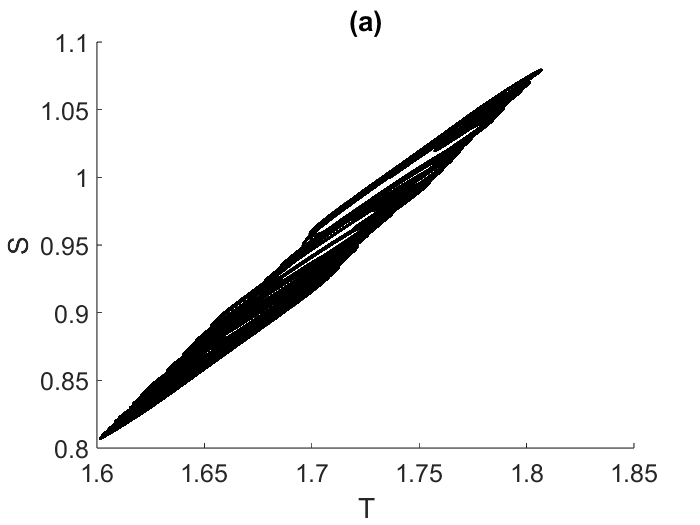}
    \includegraphics[width=0.4955\textwidth]{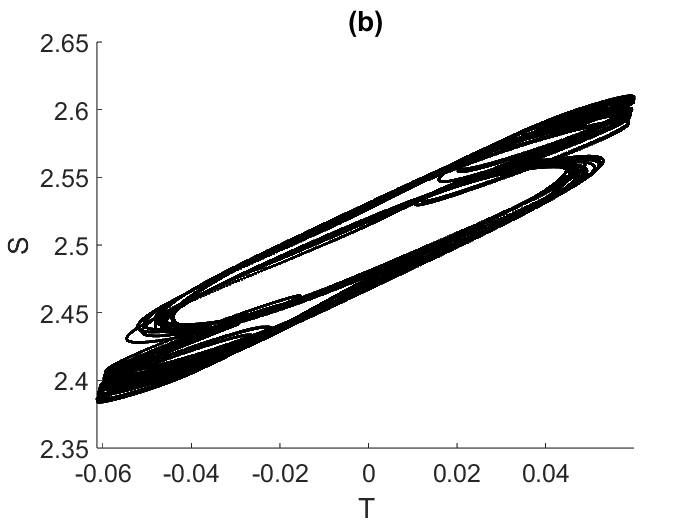}
    \includegraphics[width=0.4955\textwidth]{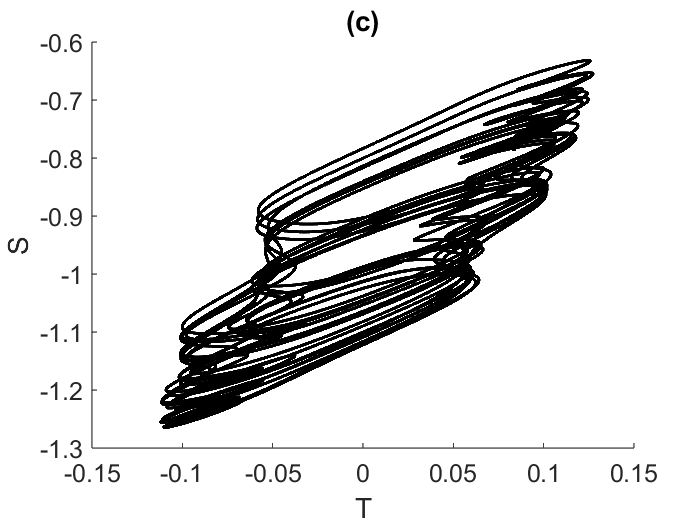}
    \includegraphics[width=0.4955\textwidth]{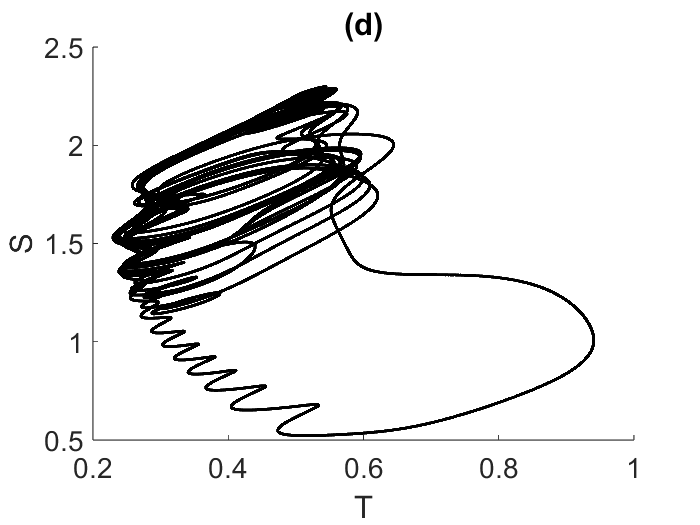}
    \caption{A selection of chaotic attractors that can occur in the LFSBM system. In (a) and (c) we illustrate two typical single-regime chaotic attractors. Figure (b) provides an example of a dual-regime chaotic attractor in which two small regimes are connected by one larger loop. Finally, in (d) we show an example of excursive behaviour for a single-regime attractor.  Parameters for each of the cases are as follows: (a) $(a,\xi,\eta,\zeta)=(0.0159,3,1,0.3)$, (b) $(a,\xi,\eta,\zeta)=(0.0226,0,1,-2.1)$, (c) $(a,\xi,\eta,\zeta)=(0.0159,3,1,0.3)$ and (d) $(a,\xi,\eta,\zeta)=(0.0541,1,-1,-2)$.}
    \label{fig:CATypes}
\end{figure}

We consider our LFSBM model defined by the system (\ref{eqn:LFSBM}). With no forcing ($a$ $=$ $0$), then the LFSBM system behaves exactly like the SBM (\ref{eqn:SBM}) and the L63 (\ref{eqn:L63Model}) models in their respective phase planes. When we introduce forcing $a$ ($> 0$), while the solutions the the Lorenz equations remain invariant, the SBM solutions change (unless we choose the initial condition of the L63 model to be the origin). When the L63 solution is chaotic, the equilibria in the Stommel phase plane become chaotic attractors. Two types are possible depending on the number of regimes; here a regime is a distinct region in phase space within which the trajectory spends an extended amount of time. Trajectories may transit between the regions but the transfer time from one to the other is much shorter than the characteristic duration of residence in a region. The possibilities for the chaotic attractor are then:
\begin{itemize}
    \item Single-Regime: These are chaotic attractors that possess only one regime in the Stommel phase plane. They are rather diverse in character and their stability properties can vary. Some examples are shown in Figure \ref{fig:CATypes} (a,c,d). A single-regime chaotic attractor can sometimes exhibit excursive behaviours where solution trajectories may temporarily deviate away from the main attractor (see Figure \ref{fig:CATypes} (d)).
    \item Dual-Regime: Chaotic attractors that possess two regimes in the Stommel phase plane are labelled dual-regime attractors. Such structures often assume a distinct shape consisting of two mini-loops augmented by one larger loop that connects them. An example of one such dual-regime attractor is shown in Figure \ref{fig:CATypes} (b).
\end{itemize}

Once the forcing reaches some critical value $a_c$, the LFSBM undergoes a crisis where the ``bistable" nature of the system changes into one of ``monostability"; this is a consequence of a change in the nature of one or more chaotic attractors. Mehra and Ramaswamy \cite{mehra1996maximal} identified three types of crises in systems that possess multiple chaotic attractors (such as the LFSBM):
\begin{itemize}
    \item Boundary Crises: A chaotic attractor is destroyed (typically as a result of a collision with a saddle) and is reduced to a chaotic transient that flows into another attractor.
    \item Interior Crises: The size of a chaotic attractor suddenly increases or decreases owing to a collision with the stable manifold of an unstable periodic orbit inside its basin of attraction.
    \item Attractor-Merging Crises: Two or more chaotic attractors simultaneously collide with the stable manifold of an unstable periodic orbit along a shared boundary in the basin of attraction. The chaotic attractors then fuse together into one attractor.
\end{itemize}
In their work Mehra and Ramaswamy use variations in the maximal Lyapunov exponent (MLE) as a guide as to whether an interior or attractor-merging crisis is likely to occur. The MLE is defined to be
\begin{equation}
    \lambda = \lim_{t\rightarrow\infty}\lim_{\delta_0\rightarrow0}\frac{1}{t}\ln\frac{|\delta(t)|}{|\delta_0|},
\end{equation}
where $\delta_0$ is a perturbation to an initial point of a trajectory and $\delta(t)$ is the evolution of that perturbation under the linearized dynamics. In our model (\ref{eqn:LFSBM}), the MLE is dictated by the chaotic Lorenz equations and thus always remains constant ($\approx 0.9057$). This implies that interior and attractor-merging crises cannot occur in the LFSBM since they require significant changes in the MLE \cite{mehra1996maximal}; however, the possibility of boundary crises cannot be ruled out.

\subsection{Chaotic Saddle Collisions}\label{sec:csc}

As noted previously, a boundary crisis typically occurs when a chaotic attractor collides with a chaotic saddle. In order to illustrate such an event, we examine the evolution of the chaotic attractors and saddle for the system (\ref{eqn:LFSBM}).  We chose the set of Stommel parameters $\xi$ $=$ $1$, $\eta$ $=$ $-1$ and $\zeta$ $=$ $-2$  which led to a chaotic transient for strong forcing. With these particular parameter values the critical forcing lies somewhere in the interval $a_{c}$ $\in$ $(0.0541, 0.0542)$. We explore the effect of increasing the forcing strength beyond $a_c$ and do so by computing basins of attraction for the chaotic attractors using the method outlined in \ref{app:BoA}.

In order to visualize the saddle, we adopted the so-called \textit{Saddle-Straddle Algorithm} \cite{battelino1988multiple,nusse2012dynamics,wagemakers2020saddle}. The algorithm was originally developed in \cite{battelino1988multiple} as a way to detect segments which belong to the saddle. We implemented the algorithm in the manner described in \cite{wagemakers2020saddle}. In brief, the algorithm works by first selecting pairs of points which straddle the basin boundary by a predetermined length $\delta$ (we used $\delta=10^{-5}$).  The points are then iterated forward under the dynamics for some chosen window and refined again to ensure that they again straddle a small segment. We then assume the midpoint of the resulting segment to be part of the chaotic saddle.

\begin{figure}
    \centering
    \includegraphics[width=0.4955\textwidth]{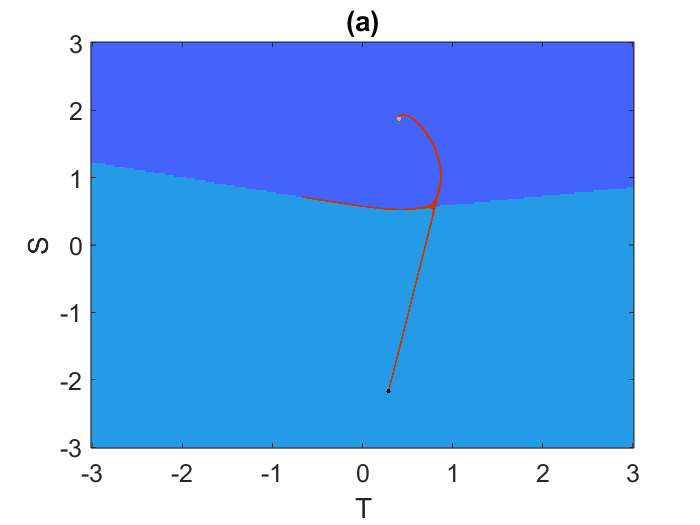}
    \includegraphics[width=0.4955\textwidth]{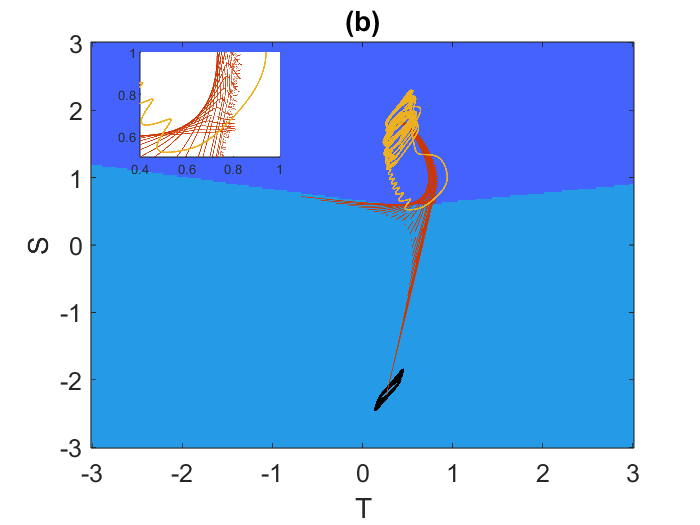}
    \includegraphics[width=0.4955\textwidth]{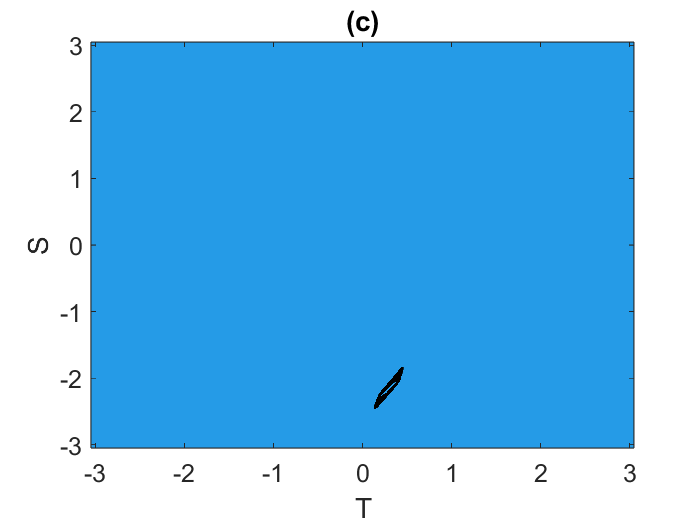}
    \includegraphics[width=0.4955\textwidth]{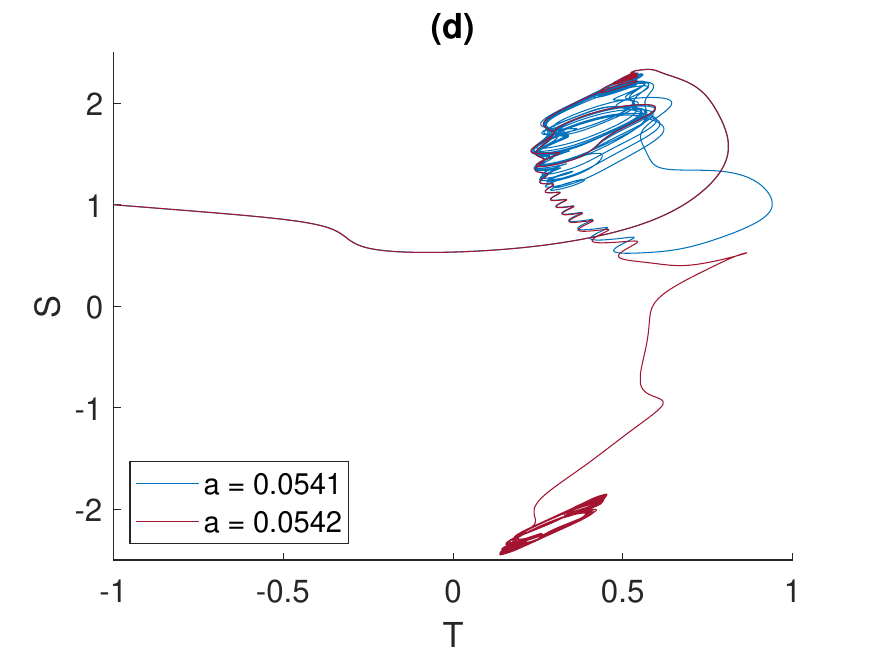}
    \caption{The basins of attraction for Stommel parameters $(\xi, \eta, \zeta)$ $=$ $(1, -1, -2)$ and Lorenz initial conditions $(x_0, y_0, z_0)$ $=$ $(-1, -1, -1)$. Three examples are presented for the forcing values $a$: (a) $0$, (b) $0.0541$ and (c) $0.0542$. The images in the top row show the thermally-driven (TH, black) and saline-driven (SA, gold) attractors and their basins of attraction in light and dark blue shadings respectively. These remain stable until the boundary crisis occurs at some value $0.0541<a< 0.0542$ when the SA attractor is destroyed (see panel c)). In (d) is an example trajectory with initial conditions $(T, S)$ $=$ $(-1, 1)$; this illustrates the chaotic transient that remains and the visible point of divergence between the two trajectories. The chaotic saddle is shown in dark orange in (a) and (b), which we show colliding with the SA attractor in (b).}
    \label{fig:BoundCrisis}
\end{figure}

Figure \ref{fig:BoundCrisis} shows the results predicted by this algorithm at a series of increasing forcing strengths.  We see that as $a$ grows,  so the saddle seems to expand in width (particularly in the more central portion of its `T'-shape). Simultaneously, the SA (saline-driven) attractor grows in phase space along the unstable manifold and approaches the stable manifold from above.  Immediately prior to the crisis ($a=0.0541$) we see that a part of the SA attractor almost touches the chaotic saddle manifold. Just after the crisis ($a=0.0542$) the trajectory which shadows the previous SA attractor diverges from that attractor near the apparent collision point.  This creates a chaotic transient which is not part of the surviving TH (thermal-driven) attractor.

The behaviour depicted in Figure \ref{fig:BoundCrisis} is particularly interesting when viewed from a finite-time standpoint. While the asymptotic behaviour after the crisis is that of convergence to the TH attractor, the system spends a long period of time elsewhere in phase space tracing the previously existing SA attractor. If this process were analyzed in isolation, this might seem to be a potentially reversible transition between the SA and TH states, whereas in actuality the crisis has already occurred and the system is destined to eventually converge to the TH attractor. Similar behaviour has previously been seen in quasi-periodically forced delay models \cite{quinn2019effects,quinn2018mid} and in systems with delayed Hopf bifurcations (see e.g. \cite{goh2022delayed} and references therein). In both the aforementioned cases, the system undergoes a bifurcation in which the stability of one attractor is lost, but if initial conditions are sufficiently close to that attractor, then the trajectory can remain nearby for an extended period of time before converging to the true attractor. Such behaviour is important to understand when considering reversible and irreversible transitions or regime shifts.

\subsection{Vanishing Basin Crisis}\label{sec:vbc}

We identify a fourth family of crises which we shall refer to as `vanishing basin'. A vanishing basin crisis is a special case of a boundary crisis.
\smallskip

\textbf{Definition}: [Vanishing Basin Crisis] Let $B_a(\omega(\textbf{\tt x}_0))$ be the basin of attraction for a chaotic attractor $\omega(\textbf{\tt x}_0)$ at some prescribed forcing strength $a$. If, for some $a$ $=$ $a_0$, we have
\begin{itemize}
    \item $B_{a_0 + \varepsilon}(\omega(\textbf{\tt x}_0))$ $\subset$ $B_{a_0}(\omega(\textbf{\tt x}_0))$ (and $B_{a_0 + \varepsilon}(\omega(\textbf{\tt x}_0))$ $\neq$ $\emptyset$) when $0$ $<$ $\varepsilon$ $<$ $A$ (where $A$ $=$ $a_c - a_0$ $>$ $0$), and
    \item  $\lim_{\varepsilon \to A^-} |B_{a_0 + \varepsilon}(\omega(\textbf{\tt x}_0))|$ $=$ $0$ (where $|\cdot|$ denotes an appropriate measure of set size),
\end{itemize}
then we say that $\omega(\textbf{\tt x}_0)$ undergoes a vanishing basin crisis. Furthermore, it loses its basin of attraction completely at some $a_c$. We remark that for the vanishing basin crisis, we only consider the behaviour of the basin of attraction in the the Stommel phase plane. We fix the initial conditions of the L63 equations for each case: $(x, y, z)$ $=$ $(x_0, y_0, z_0)$ when $t=0$.
\smallskip

\begin{figure}
    \centering
    \includegraphics[width=0.4955\textwidth]{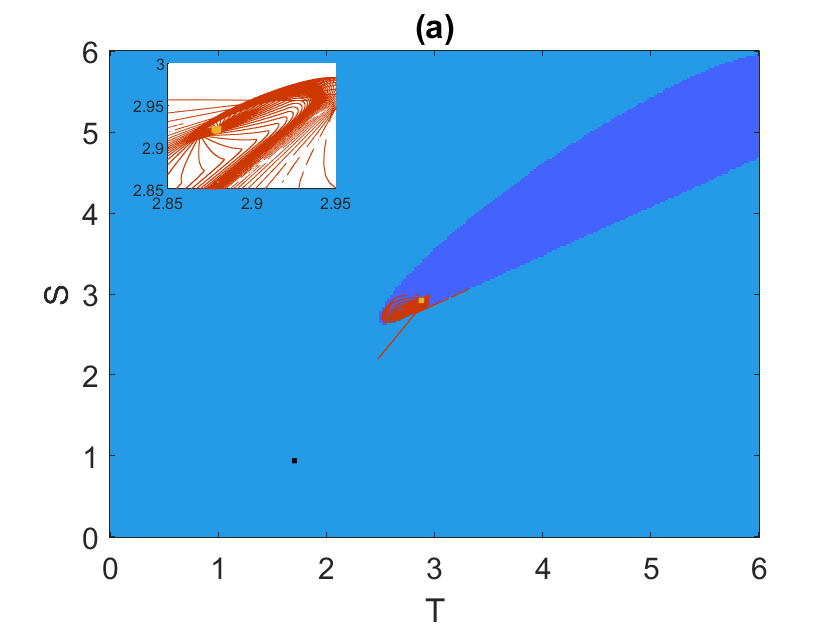}
    \includegraphics[width=0.4955\textwidth]{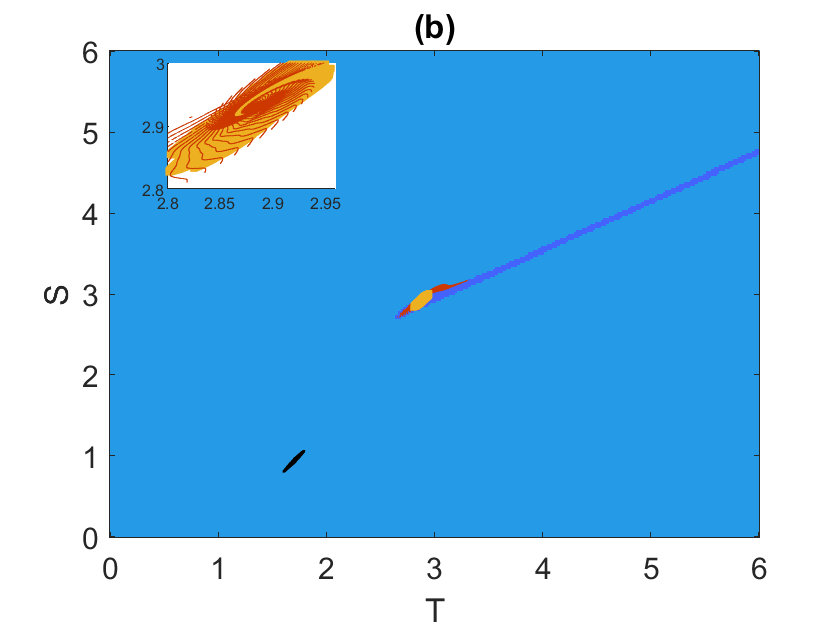}
    \includegraphics[width=0.4955\textwidth]{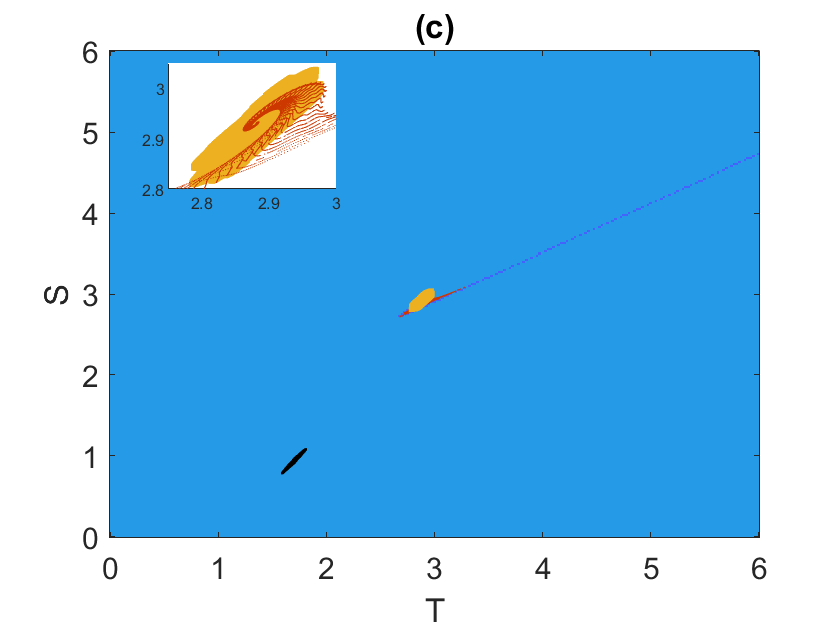}
    \includegraphics[width=0.4955\textwidth]{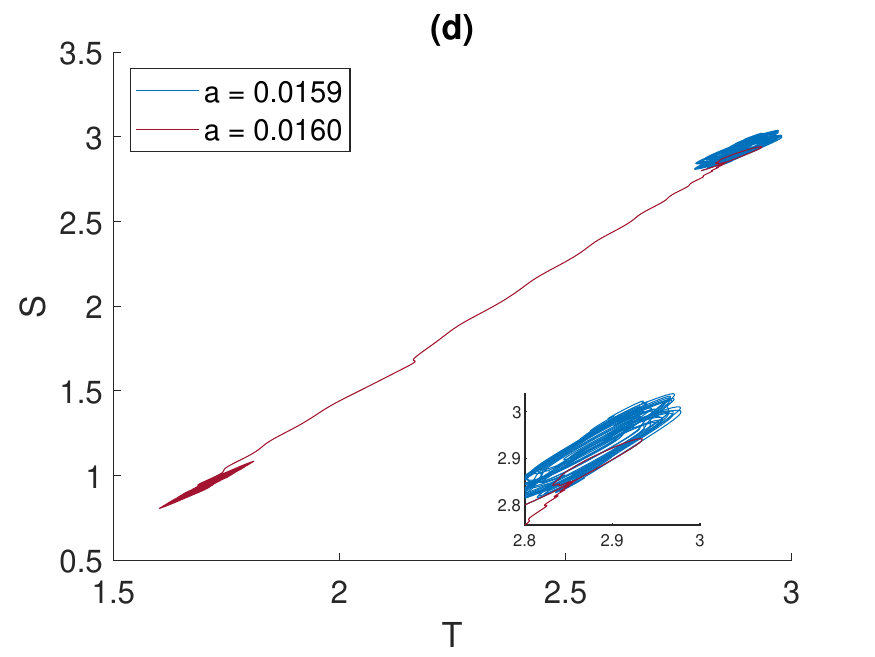}
    \caption{The basins of attraction for the system with Stommel parameters $(\xi, \eta, \zeta)$ $=$ $(3, 1, 0.3)$ and Lorenz initial conditions $(x_0, y_0, z_0)$ $=$ $(1, 1, 1)$. Three examples are shown with values of forcing $a=$ (a) $0$, (b) $0.0135$ and (c) $0.0159$. The images in the top row show the thermally-driven (TH, black) and saline-driven (SA, gold) attractors and their basins of attraction in light and dark blue shadings respectively. As we increase $a$, we see that the basin of attraction for the SA attractor shrinks until it eventually disappears altogether beyond $a$ $=$ $0.0159$. This is reflected in (d), where an example solution trajectory $(T_0, S_0)$ $=$ $(2.8, 2.8)$ fails to find the SA attractor when $a$ $=$ $0.0160$. We include the chaotic saddle (colored dark orange) in our basin calculations, showing how the chaotic saddle wraps up around itself under forcing (with extra detail in the insets).}
    \label{fig:VBCrisis}
\end{figure}

Unlike other crises, the vanishing basin crisis does not occur instantaneously. Rather, the crisis develops with an increasing forcing strength, starting from some $a_0$ before the attractor eventually loses its stability at a critical value $a_c$ $(>a_0)$. As the forcing increases from $a_0$ to $a_c$, it can be shown that the size of the basin of attraction shrinks. At the completion of the crisis, the chaotic attractor is not destroyed. Instead, it simply loses its basin of attraction and becomes a ghost attractor (a designation coined by Belykh et al. \cite{belykh2013multistable}). We provide an example of a vanishing basin crisis in Figure \ref{fig:VBCrisis}. To the best of our knowledge, this particular type of crisis has not been observed in previous literature on chaotic attractors. We note that a vanishing basin was observed in a similar box model for the ocean (without chaotic forcing) which led to interesting rate-induced dynamics \cite{alkhayuon2019}.

\subsection{Beyond Critical Forcing}\label{sec:beyond_crit}

\begin{figure}
    \centering
    \includegraphics[width=0.4955\textwidth]{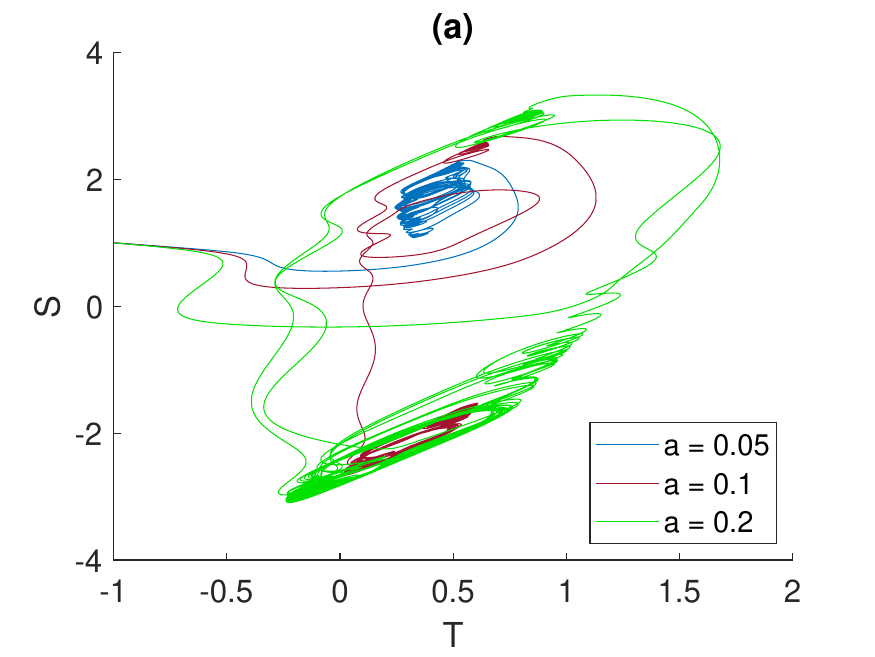}
    \includegraphics[width=0.4955\textwidth]{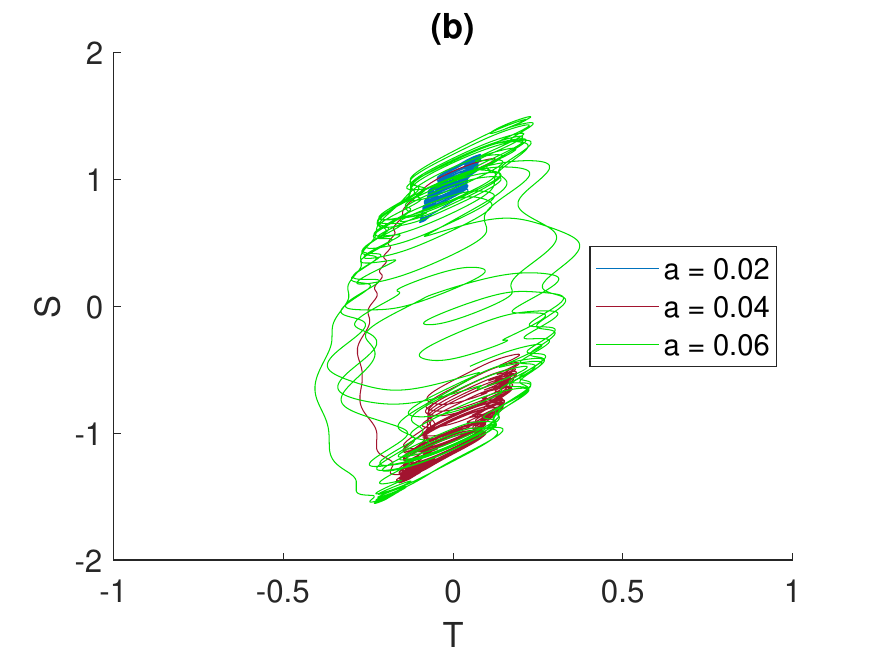}
    \caption{Examples of a chaotic transient merge (a) and a ghost attractor re-appearance (b). The parameter values for (a) are identical to those used for figure \ref{fig:BoundCrisis}, while for (b) the parameter values are $(\xi, \eta, \zeta)$ $=$ $(0, 0, -1)$ with initial condition $(x_0, y_0, z_0, T_0, S_0)$ $=$ $(1, 1, 1, 0, 1)$. The forcing $a$ is increased such that the trajectories evolve through pre-crisis, post-crisis and post-merge regimes; these are shown blue, red and green respectively.}
    \label{fig:GA+CTMerge}
\end{figure}

As the forcing increases beyond $a_c$, the general regime pattern of the surviving attractor remains the same as that which existed shortly after the crisis. The attractor continues to develop until a second critical forcing level $a_r$ is reached whereupon, though the system itself remains monostable, the chaotic attractor undergoes a significant structural change. The two types of structural changes are as follows:
\begin{itemize}
    \item If the system suffers a typical boundary crisis, then the resulting chaotic transient persists in that region of phase space before merging with the remaining chaotic attractor at the forcing value $a_r$ (see figure \ref{fig:GA+CTMerge} (a)).
    \item If the system undergoes a vanishing basin crisis, then the attractor that loses its associated basin simply vanishes from view, becoming a ghost attractor. Solution trajectories attempt (and fail) to locate it before entering the other attractor until the forcing reaches the value $a_r$. Once $a$ $=$ $a_r$, the attractor with the vanishing basin re-appears and merges with the persisting chaotic attractor (Figure \ref{fig:GA+CTMerge} (b)).
\end{itemize}

\section{Finite-time Analysis\label{sec:results}}

In this section we employ a method to investigate the local behaviour in time along a given trajectory. For a boundary crisis, there exists a point along a trajectory such that the attractor collides with a chaotic saddle. As a saddle consists of both attracting and diverging directions, we expect to see evidence of divergence when approaching the saddle. Such divergence of nearby trajectories is captured by the Lyapunov exponents, which are defined as the asymptotic growth and decay rates of small perturbations to a trajectory. We consider a finite-time version of the Lyapunov exponents, or FTLEs, calculated using a sliding window along the trajectory. This confines the growth or decay behaviour to the corresponding section of the trajectory. We expect that trajectories approaching a saddle would have a change in the FTLEs indicating that they are approaching an area of phase space with differing growth and decay characteristics. The calculation of Lyapunov exponents and FTLEs is outlined in \ref{app:NumMeth}.

We observe how the FTLEs change over a given solution trajectory for a given set of initial conditions and parameters. We calculate five FTLEs at any given point along the trajectory, three of which are associated with the Lorenz attractor forcing, and the other two with the SBM response. For our finite-time analysis, we set the step size for our FTLEs (and the ODE solver we use for the study) to be $\Delta t$ $=$ ${1}/{400}$, and chose to calculate across the range $t$ $\in$ $[0, 100]$, giving us 40,000 time steps in total. Any time step can be implemented, however. the detailed quantitative properties of the chaotic signal induced will be sensitive to the time-step and may slightly change the level of forcing required to induce a crisis. We also remark at this point that the length of the FTLE window we select for the study will give different results in terms of FTLE behaviour; smaller FTLE windows tend to give more variance in the FTLEs, while longer windows filter out this high-frequency variation. We set our default FTLE window to 400 time steps (one time unit) and only discuss the implications of using other FTLE windows. The five FTLEs we derive for the LFSBM are referred to by their asymptotic counterpart. From highest to lowest these are: the unstable Lorenz FTLE, the neutral Lorenz FTLE, the first Stommel FTLE, the second Stommel FTLE and the stable Lorenz FTLE. These designations are determined by comparing their time averages along the trajectory to the asymptotic exponents of the uncoupled models. The values of these FTLEs are not expected to be their asymptotic values, but given the system is ergodic, their mean over a sufficient window should be near to the corresponding asymptotic values.

\begin{figure}
    \centering
    \includegraphics[width=0.4955\textwidth]{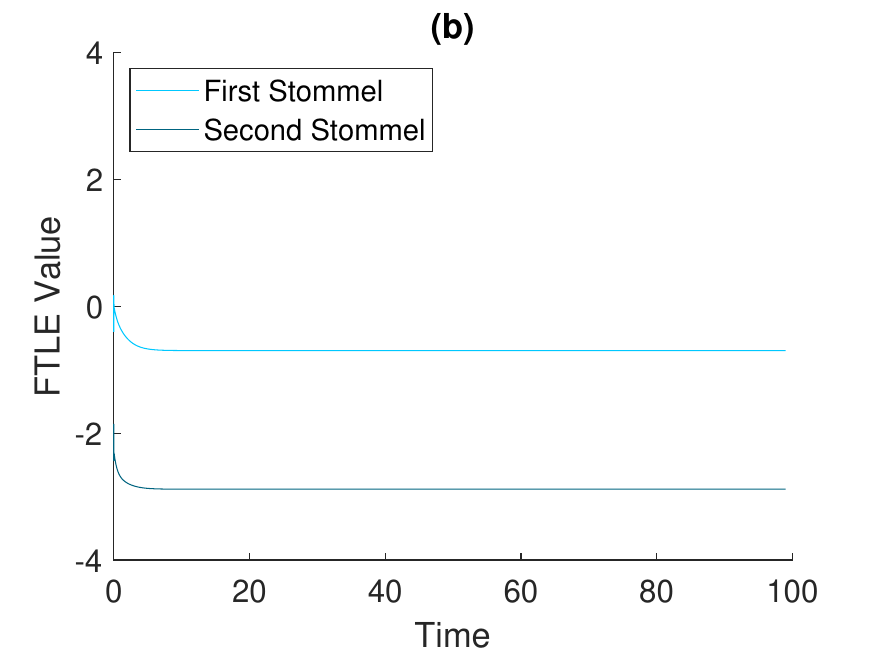}
    \includegraphics[width=0.4955\textwidth]{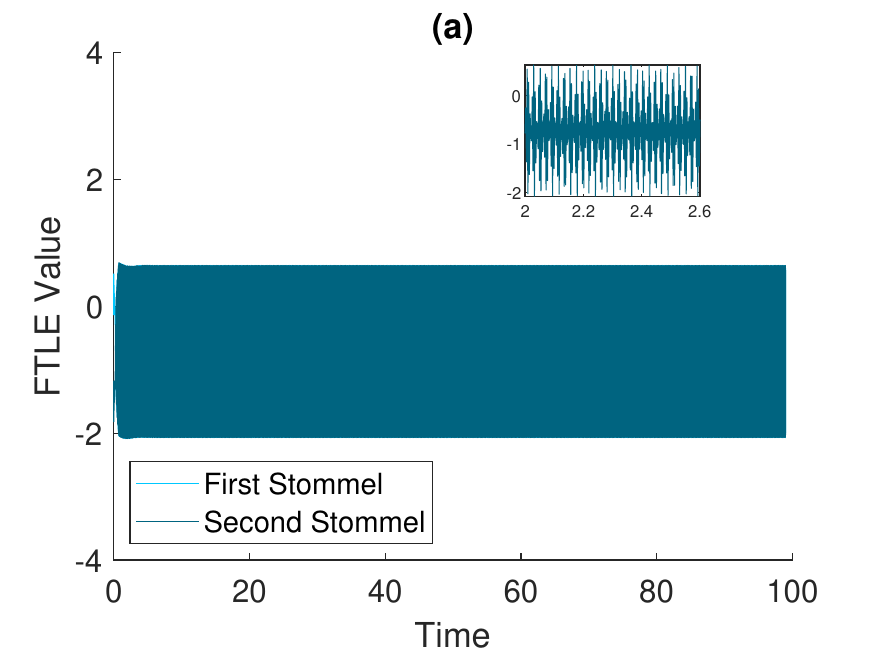}
    \caption{A demonstration of the main differences between the two Stommel FTLE behaviours in the absence of forcing ($a$ $=$ $0$). In (a), we have the distinct separation between the two FTLEs. In (b), we have a rapid oscillatory behaviour. The inset provides a zoomed-in view of the behaviour over a sufficiently small time frame.}
    \label{fig:DSvsROComp}
\end{figure}

We first consider the Stommel FTLEs in the unforced case ($a$ $=$ $0$). For a bistable system with two stable equilibria in differing regions (one in the TH region, the other in the SA region of the phase plane) the behaviour of these FTLEs will largely depend on the nature of the equilibrium in the region of the phase plane as predicted by performing Hartmann linearisation on (\ref{eqn:SBM}). If the equilibrium in the region is a stable node, then the Stommel FTLEs will converge towards the two eigenvalues of the stable node (this does not vary with different FTLE window lengths). We refer to this natural behaviour of the Stommel FTLEs as one of distinct separation. If the equilibrium in the region is instead a stable focus, then the Stommel FTLEs will oscillate with the frequency of the imaginary part of the eigenvalues. This oscillation will be centred around the value of the real part of the eigenvalues and have an amplitude that is dependent on the length of the FTLE window. We found empirically that a longer FTLE window gives an amplitude tending towards zero, while a shorter FTLE window will see the amplitude approach the imaginary part of the eigenvalues. We refer to this natural behaviour of the Stommel FTLEs as a rapid oscillation. We illustrate the two natural behaviours in Figure \ref{fig:DSvsROComp}.

\begin{figure}
    \centering
    \includegraphics[width=0.4955\textwidth]{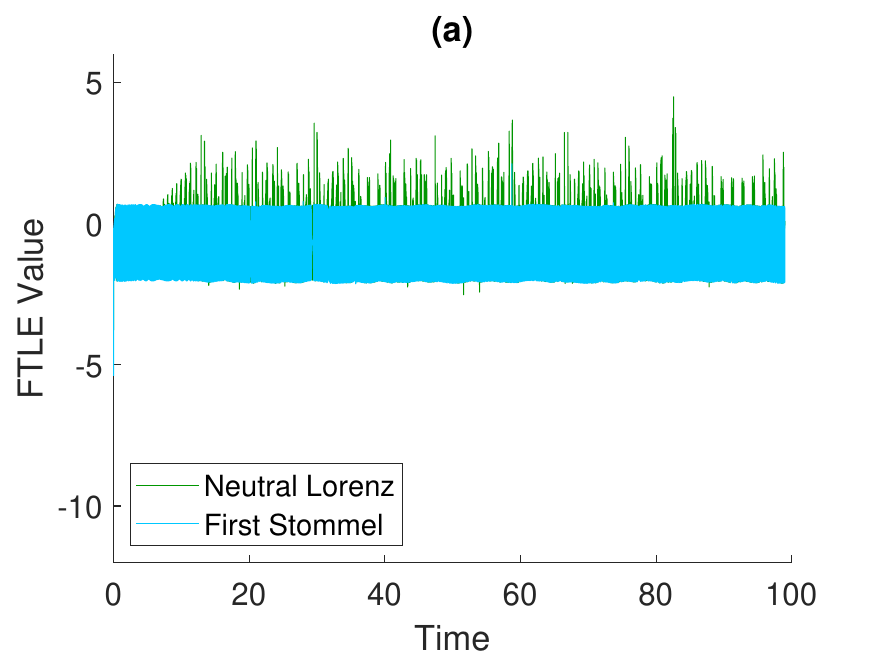}
    \includegraphics[width=0.4955\textwidth]{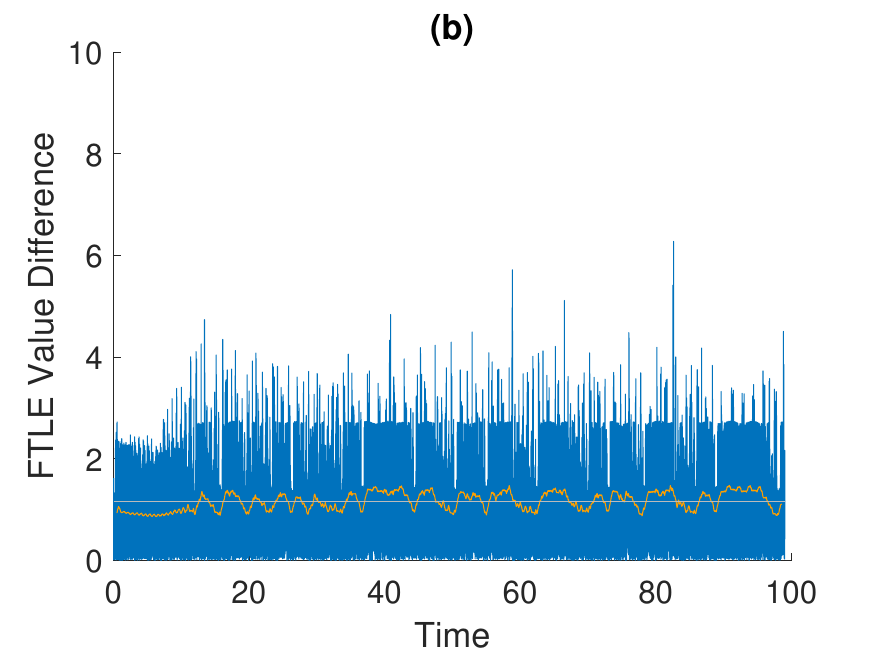}
    \includegraphics[width=0.4955\textwidth]{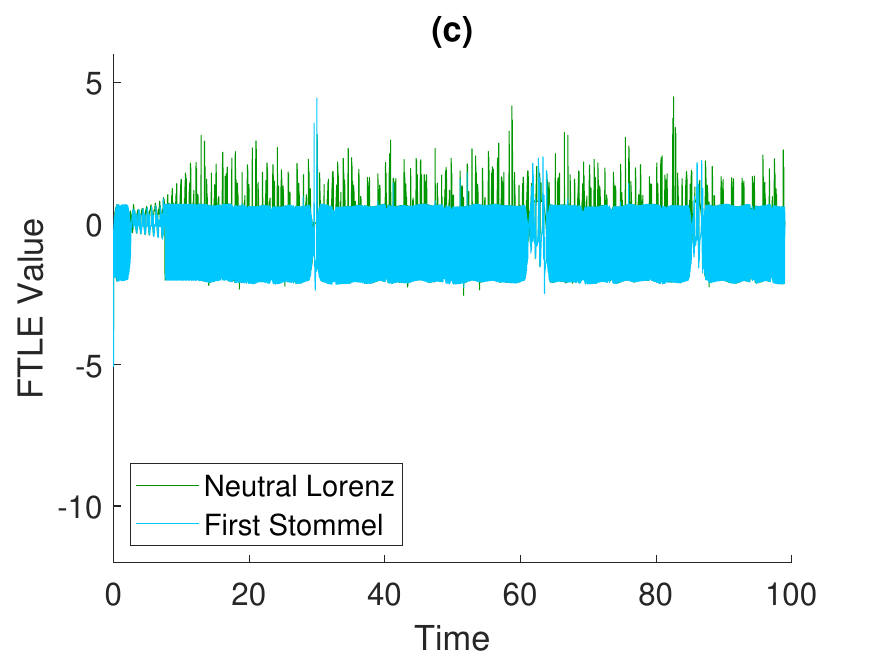}
    \includegraphics[width=0.4955\textwidth]{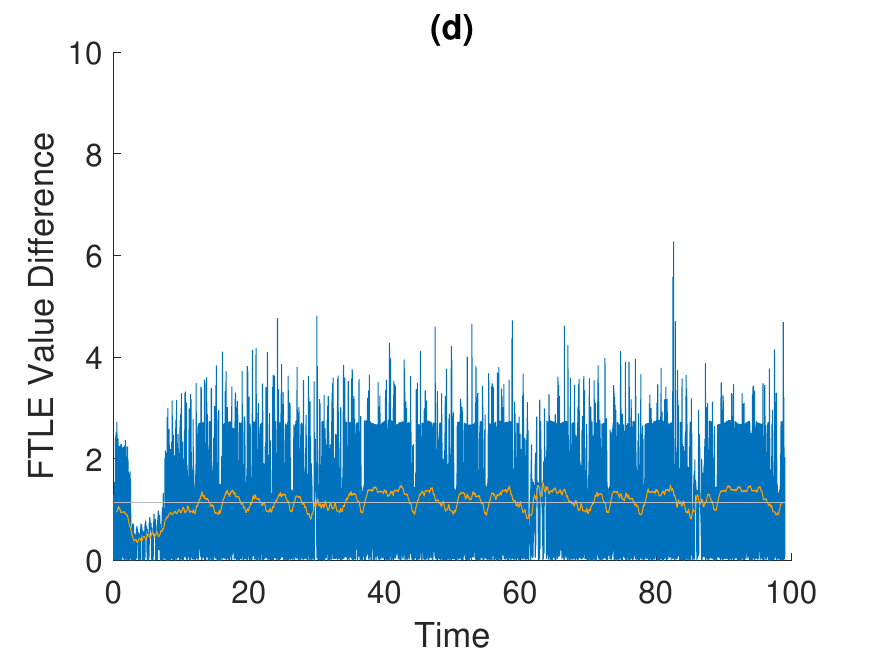}
    \includegraphics[width=0.4955\textwidth]{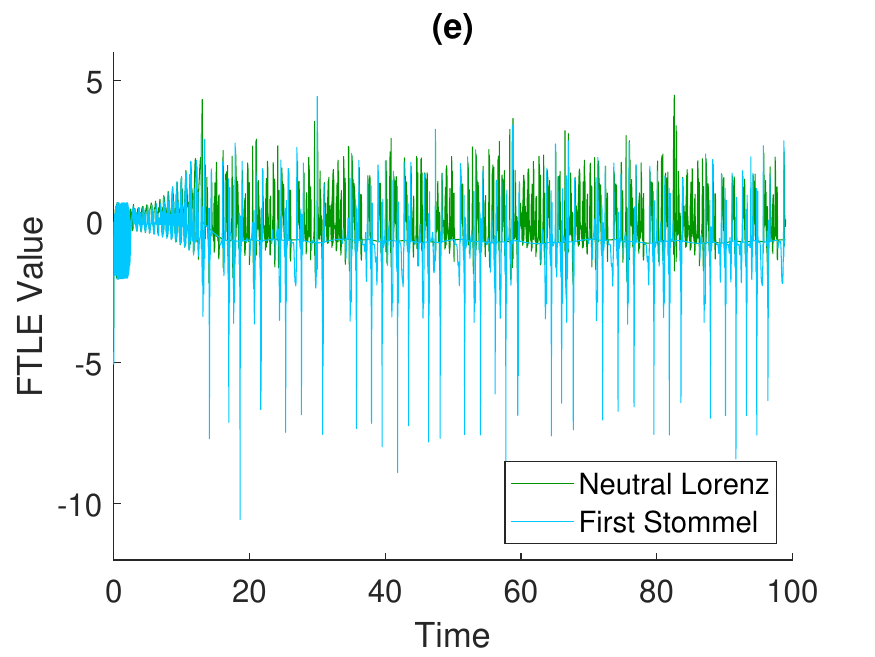}
    \includegraphics[width=0.4955\textwidth]{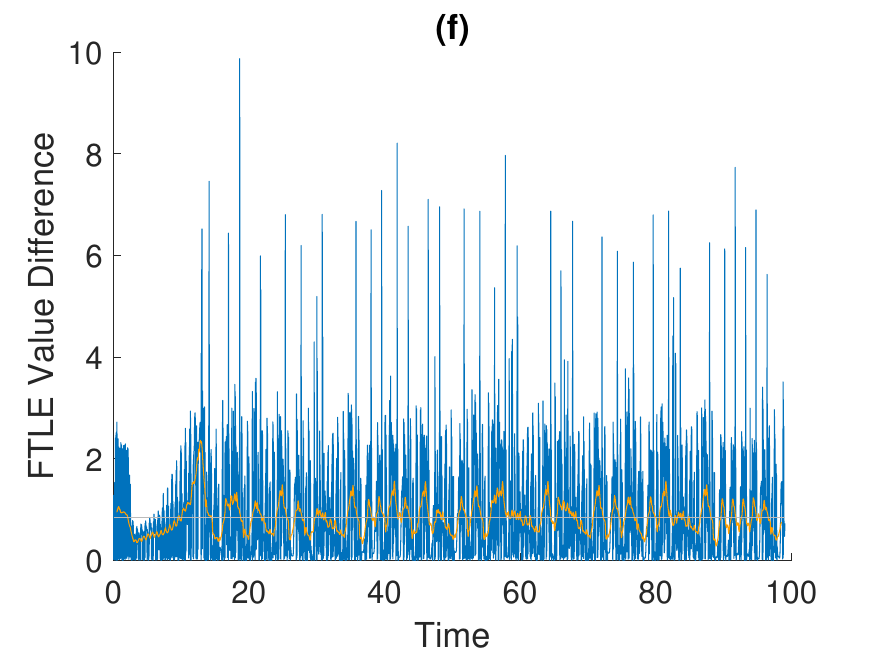}
    \caption{Illustration of how the neutral Lorenz and first Stommel FTLEs approach under forcing. Here the parameters $(\xi, \eta, \zeta)$ $=$ $(3, 1, 0.3)$ and the initial conditions $(x, y, z, T, S)$ $=$ $(1, 1, 1, 2.8, 2.8)$. We consider three forcing levels $a$; 0.01 (top row), $0.0159$ (middle row) and $0.0160$ (bottom row). In the left-hand column we observe how the first Stommel exponent starts to behave increasingly like a Lorenz FTLE as the crisis point is first approached and then exceeded (for these parameter choices the crisis occurs at a forcing value in the interval (0.0159, 0.0160)). This behaviour is confirmed by an examination of the distance between the two exponents and this quantity is plotted in the right-hand column. The orange line denotes the rolling FTLE average distance over one time unit while the grey line denotes the average distance over the full time period.}
    \label{fig:0.01F+0.0159F+0.016FComp}
\end{figure}

As forcing is gradually introduced into the model, the behaviour of the Stommel FTLEs begins to change. With the aforementioned window choice, the values of the Stommel FTLEs start to exhibit spikes at certain points along the trajectory. We remark that these spikes are smoothed out if longer window lengths are used and this leads to a loss of local information. If the trajectory of a solution happens to cross regions (\emph{e.g.} from TH to SA, temporarily or otherwise), the FTLEs will typically adopt the behaviour associated with the region in which the trajectory resides. This general behaviour persists as the forcing strengthens until it nears the crisis point. Around this stage, we notice some significant differences in the behaviour of the Stommel FTLEs (though this will vary from case to case). One notable change relates to FTLE convergence, particularly between the first Stommel and neutral Lorenz FTLEs. We present some typical results in Figure \ref{fig:0.01F+0.0159F+0.016FComp}.

\begin{figure}
    \centering
    \includegraphics[width=0.4955\textwidth]{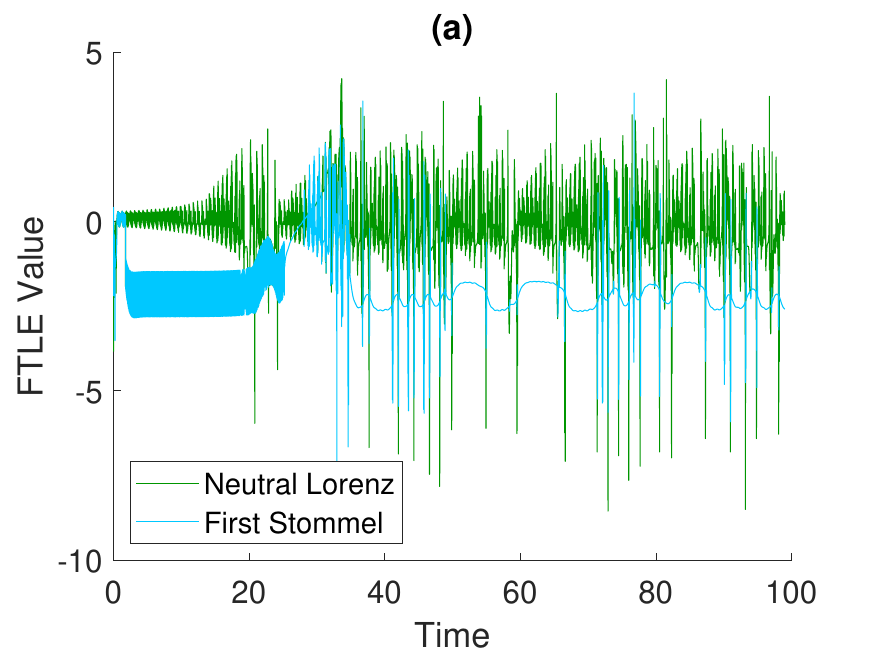}
    \includegraphics[width=0.4955\textwidth]{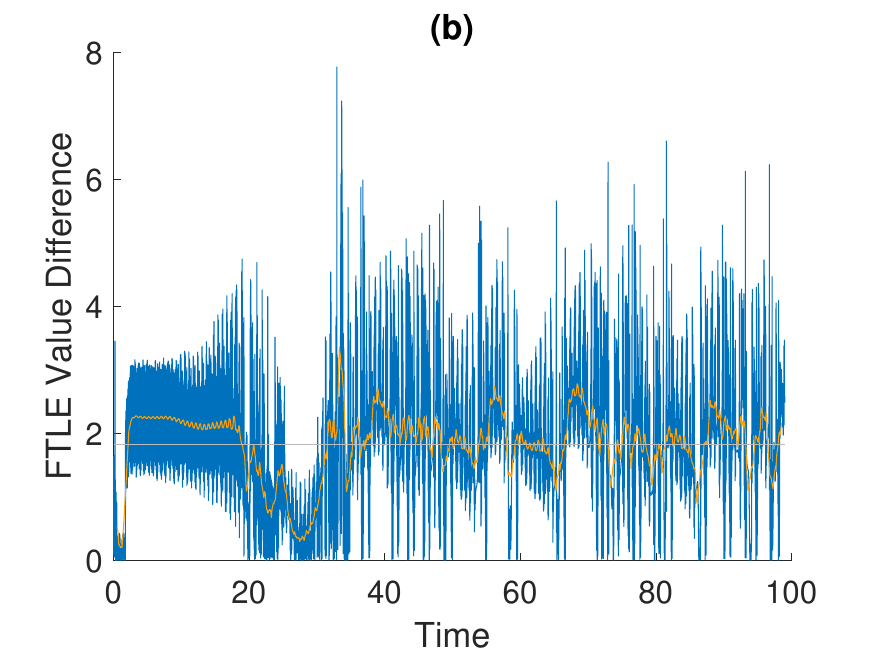}
    \includegraphics[width=0.4955\textwidth]{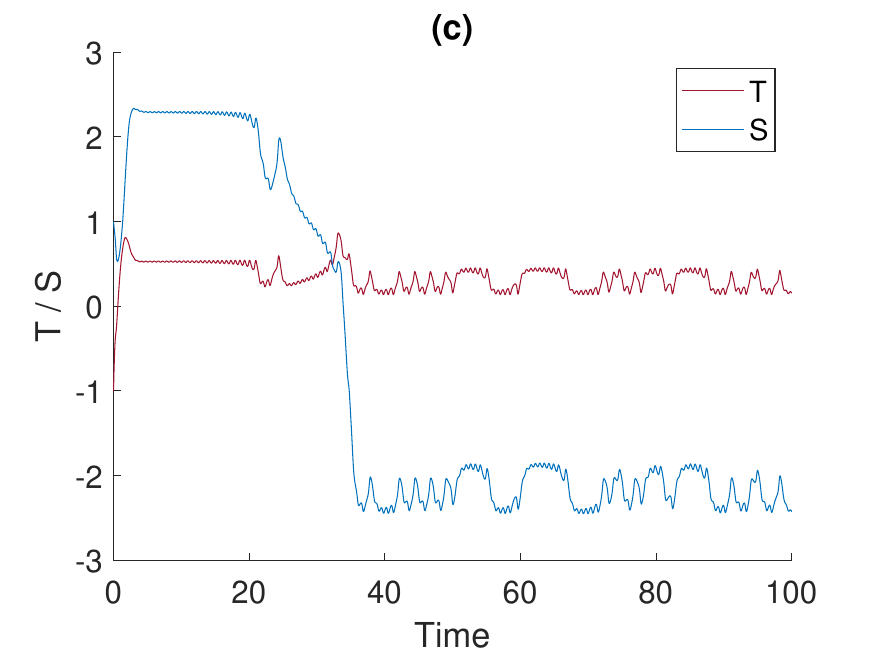}
    \includegraphics[width=0.4955\textwidth]{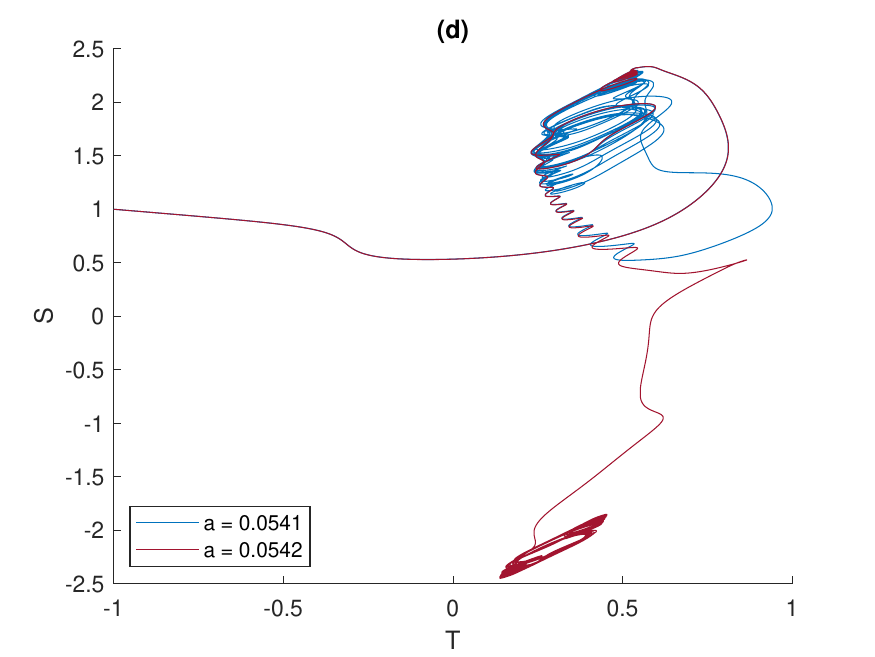}
    \caption{Results of finite-time analysis for parameters $(\xi, \eta, \zeta)$ $=$ $(1, -1, -2)$ and initial condition $(x, y, z, T, S)$ $=$ $(-1, -1, -1, -1, 1)$. The forcing level $a$ $=$ 0.0542 which is is in the post-crisis regime (see Figure \ref{fig:BoundCrisis}). In (a) we show the neutral Lorenz and first Stommel FTLEs with the two becoming visibly close in the $t$ $\in$ $[20, 40]$ range. This is confirmed by the form of the FTLE distances shown in (b), where it is seen that there is an interval over which the distance is small (both the raw values in blue, rolling average over one time unit in orange, and average over full time period in grey). After this interval the FTLE distance returns to more usual values. The results in (c) demonstrate conclusively that the time over which the FTLE distance is small coincides with the stage at which the trajectory crosses from the SA region to the TH region. Finally, in (d) are shown trajectories at two values of the forcing either side of its value when the crisis occurs.}
    \label{fig:1-1-2Comp}
\end{figure}

To assess the convergence between the neutral Lorenz and first Stommel FTLEs, we use a simple absolute distance metric defined by
\begin{equation}
    d(x_1, x_2) := |x_1 - x_2|
    \label{eqn:1Norm}
\end{equation}
for scalars $x_1$ and $x_2$ (which represent the two FTLEs at some given time). We measure the gap between the two FTLEs at a given instant in time, taking lower distances to be indicative of a stronger alignment. Using this metric, we can make two general observations regarding FTLE behaviour around crises. We see that for the vanishing basin case, the first Stommel FTLE and neutral Lorenz FTLE show decreasing distance as the system approaches the crisis value of the forcing. Figure \ref{fig:0.01F+0.0159F+0.016FComp} demonstrates this through increasing instances of the distance approaching zero as the forcing increases to the critical crisis level. We only see a small decrease in the mean distance along the trajectory, but this is likely impacted by the large transient excursions of the Stommel exponent. We also found that just before a general boundary crisis, the distance is smallest when the trajectory is at a point such that only a slight increase in the forcing strength causes an escape to the other attractor (Figure \ref{fig:1-1-2Comp}). It is possible that for the vanishing basin crisis, there are in fact many critical points and the overall decrease in distance is a reflection of this property. We also note that convergence of FTLE values in the critical region persists even after the forcing surpasses the critical value. This convergence is more evident in cases where the attractor that loses stability resides in the SA region, but we note that this is still present in examples involving crises in the TH attractor. The general convergence between the neutral Lorenz and first Stommel FTLEs across the entirety of a trajectory before and after a crisis is dependent on both the system and the initial conditions.

The alignment of FTLE behaviour, particularly into the neutral subspace, has been previously observed in similar conceptual chaotic models when they undergo a transition between regions of the attractor, as well as in models of regime shifts in atmospheric flows \cite{quinn2020,quinn2021,maiocchi2024}. In general, however, more useful measures for alignment strength already exist in regards to Lyapunov vectors, which not only help characterize the dynamics of a system \cite{ginelli2007characterizing}, but strong levels of alignment can be used to predict chaotic transitions (e.g. \cite{beims2016alignment}, \cite{sharafi2017critical}). 

\section{More general considerations\label{sec:generality}}

We consider the possible wider applicability of some of our findings by studying the properties of crises and finite-time analysis as they relate to the SBM forced by other `strange' attractors. Several candidate models were considered, but after extensive testing, we restrict our discussion to the R\"ossler, the Four-wing and the Halvorsen attractors. In particular, for the R\"ossler-forced model, we define
\begin{equation}
    \begin{array}{lcl}
        \dot{x}(t) & = & -(y + z), \\
        \dot{y}(t) & = & x + by, \\
        \dot{z}(t) & = & c + z(x - d), \\
        \dot{T}(t) & = & \xi + ax - T(1 + |T - S|), \\
        \dot{S}(t) & = & \eta + ax - S(\zeta + |T - S|),
    \end{array}
    \label{eqn:RFSBM}
\end{equation}
in which the parameters $(b, c, d)$ $=$ $(0.2, 0.2, 5.7)$. This combination is known to generate a chaotic signal \cite{letellier2006rossler}. For the Four-wing model, we defined
\begin{equation}
    \begin{array}{lcl}
        \dot{x}(t) & = & bx + yz, \\
        \dot{y}(t) & = & cx + dy - xz, \\
        \dot{z}(t) & = & -(z + xy), \\
        \dot{T}(t) & = & \xi + ax - T(1 + |T - S|), \\
        \dot{S}(t) & = & \eta + ax - S(\zeta + |T - S|),
    \end{array}
    \label{eqn:FWFSBM}
\end{equation}
with $(b, c, d)$ $=$ $(0.2, 0.02, -0.4)$, see \cite{wang20093}. Lastly, for the Halvorsen model:
\begin{equation}
    \begin{array}{lcl}
        \dot{x}(t) & = & -(bx + 4y + 4z + y^2), \\
        \dot{y}(t) & = & -(by + 4z + 4x + z^2), \\
        \dot{z}(t) & = & -(bz + 4x + 4y + x^2), \\
        \dot{T}(t) & = & \xi + ax - T(1 + |T - S|), \\
        \dot{S}(t) & = & \eta + ax - S(\zeta + |T - S|),
    \end{array}
    \label{eqn:HFSBM}
\end{equation}
where $b$ $=$ $1.27$,  as in \cite{vaidyanathan2016adaptive}. 

\begin{figure}
    \centering
    \includegraphics[width=0.4955\textwidth]{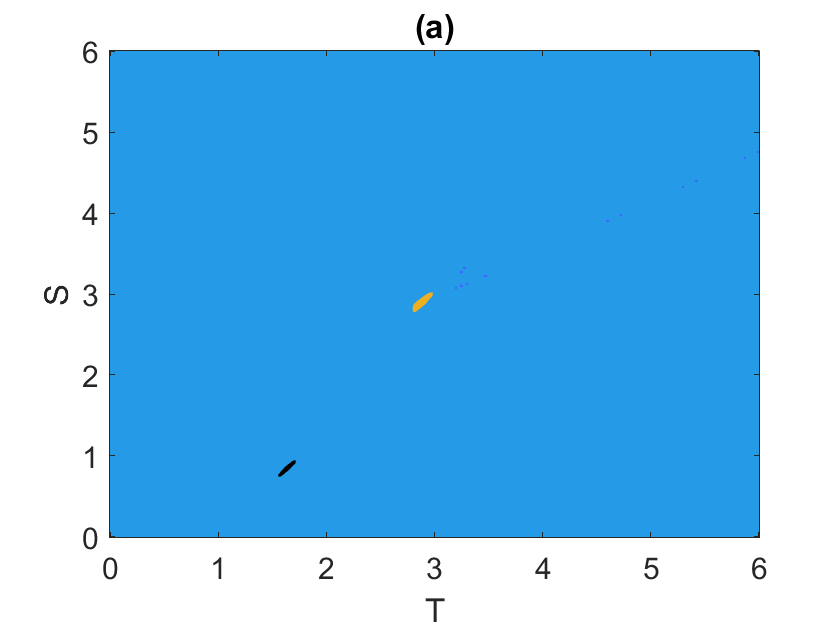}
    \includegraphics[width=0.4955\textwidth]{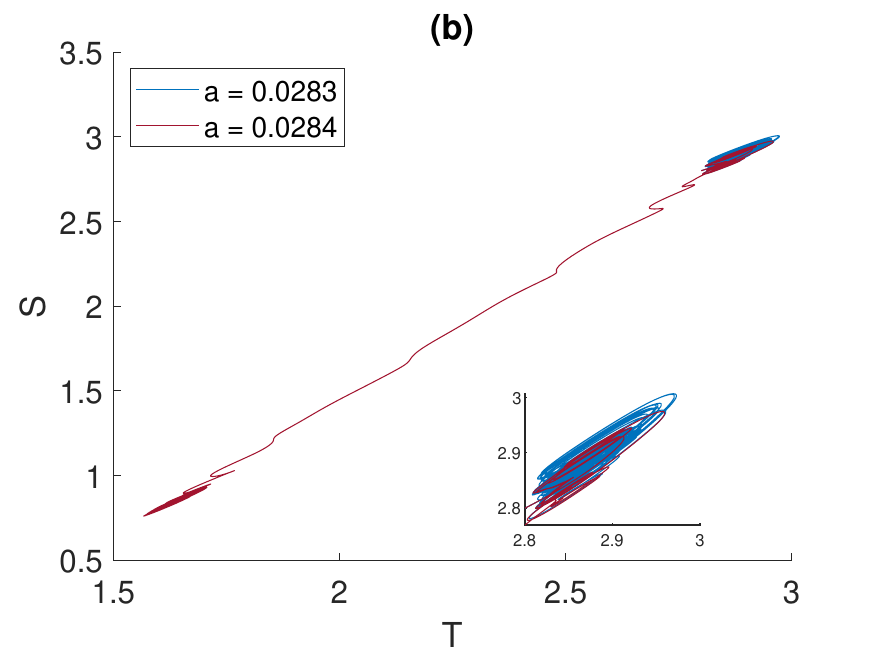}
    \includegraphics[width=0.4955\textwidth]{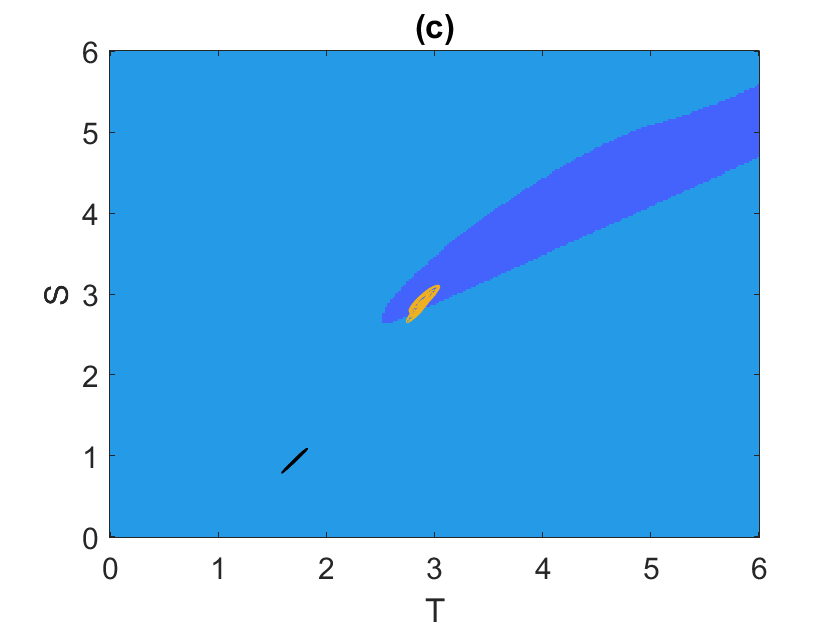}
    \includegraphics[width=0.4955\textwidth]{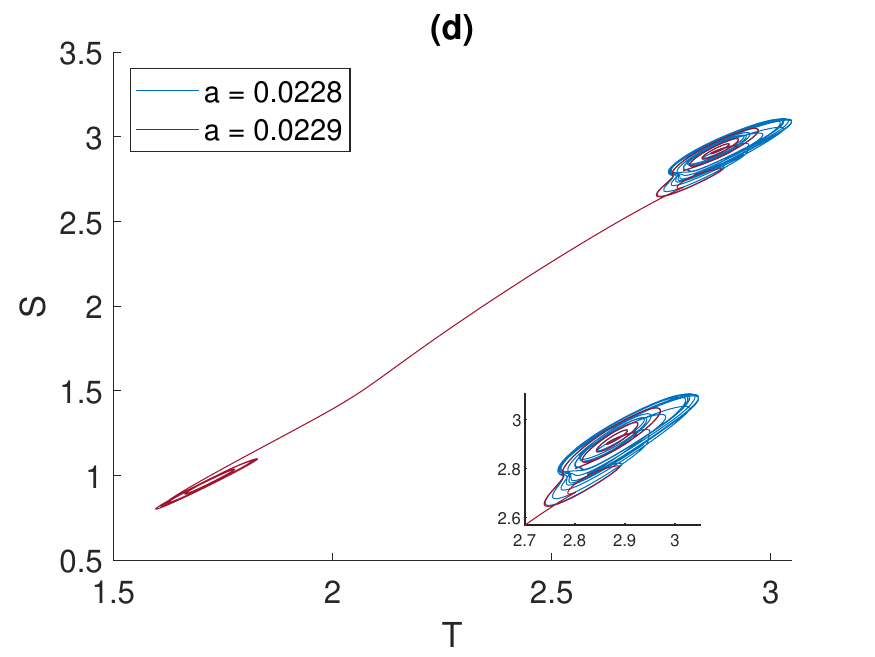}
    \includegraphics[width=0.4955\textwidth]{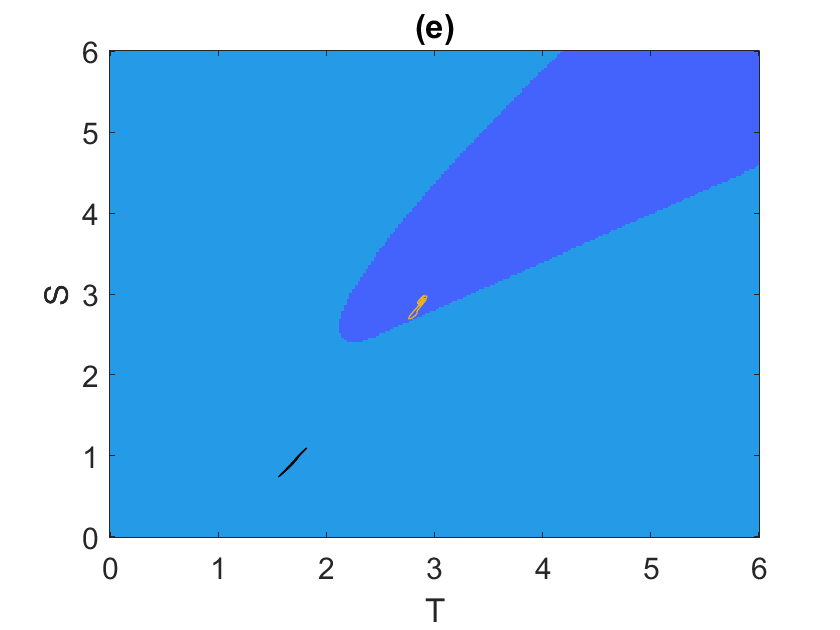}
    \includegraphics[width=0.4955\textwidth]{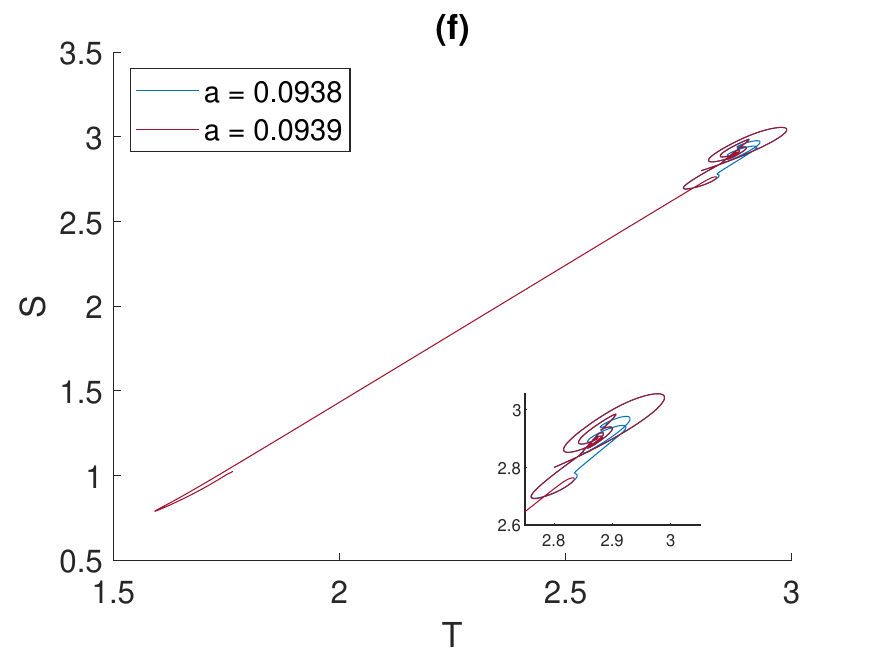}
    \caption{Crises (with chaotic saddles omitted) for the Stommel parameters, $(\xi, \eta, \zeta)$ $=$ $(3, 1, 0.3)$ with alternative chaotic forcing to the SBM: the Halvorsen model (top row);  with the R\"ossler model (middle row) and with the four-wind model (bottom row).  The images in the left column show the thermally-driven (TH, black) and saline-driven (SA, gold) attractors with their basins of attraction in light and dark blue shadings respectively. The images in the right column illustrate the respective trajectory behaviour just before and after a crises (for Stommel initial conditions $(T, S)$ $=$ $(2.8, 2.8)$). We find that the Halvorsen-forced model undergoes a vanishing basin crisis ($a$ $=$ $0.0283$), while the R\"ossler- and Four-wing-forced models exhibit a typical boundary crisis ($a$ $=$ $0.0228$ and $a$ $=$ $0.0939$ respectively).}
    \label{fig:3103BasinComp}
\end{figure}

We subject these systems to the same procedures as were applied to the LFSBM. Vanishing basin crises could be observed in the Halvorsen-forced model (\ref{eqn:HFSBM}) with similar sets of Stommel parameters as in the LFSBM. Other strange attractors tested did not seem to undergo such a crisis. We note that when $(\xi, \eta, \zeta)$ $=$ $(3, 1, 0.3)$ the R\"ossler-forced (\ref{eqn:RFSBM}) and Four-wing-forced models (\ref{eqn:FWFSBM}) possess SA attractors whose basins decrease with forcing, but undergo a regular boundary crisis when $a$ reaches the relevant critical forcing levels. These results are summarized in Figure \ref{fig:3103BasinComp}. The underlying reason why the Lorenz and Halvorsen models exhibit vanishing basin crises while the R\"ossler and Four-wing models do not is unclear. This is a possible topic for future research.

\begin{figure}
    \centering
    \includegraphics[width=0.4955\textwidth]{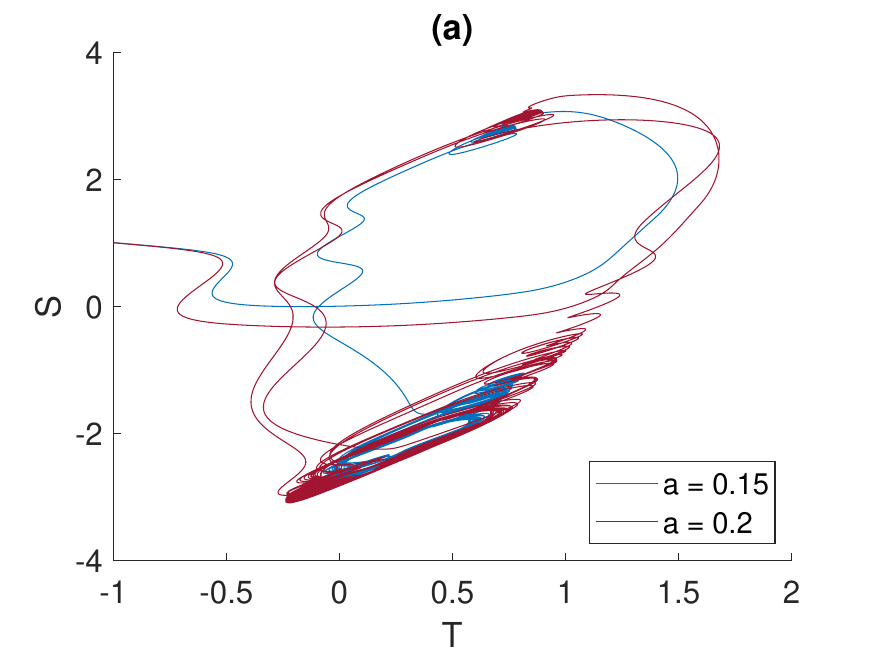}
    \includegraphics[width=0.4955\textwidth]{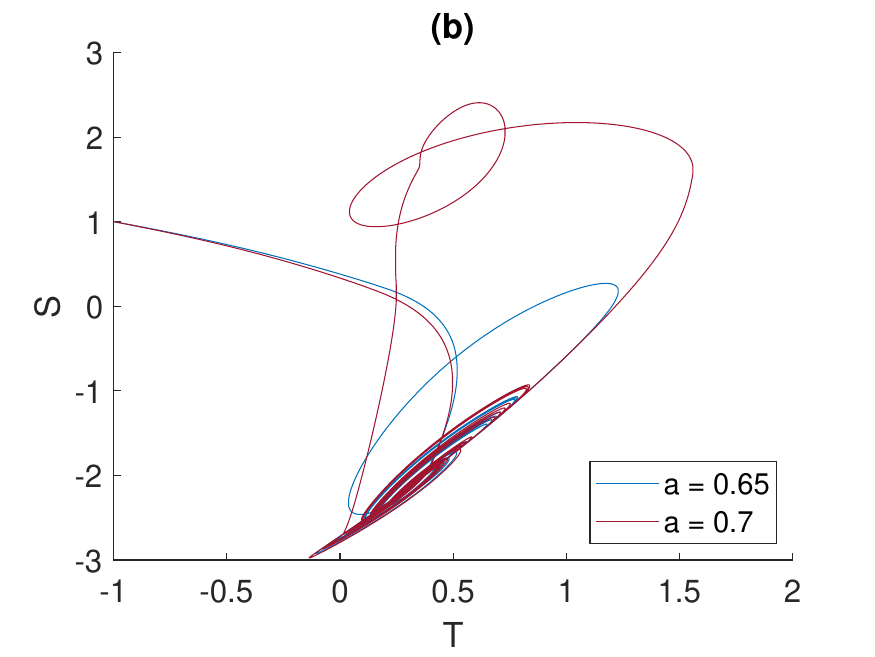}
    \caption{Chaotic transient merges under significant amounts of forcing in (a) the Lorenz-forced and (b) the Four-wing models. These calculations were performed with parameters $(\xi, \eta, \zeta)$ $=$ $(1, -1, -2)$ and $(x, y, z, T, S)$ $=$ $(-1, -1, -1, -1, 1)$. For the Four-wing model, we extend the time interval to $[0, 1000]$ to show the merging event.}
    \label{fig:1-1-2TMergeComp}
\end{figure}

These findings using other strange attractors also suggest a possible link between the number of regimes in the chaotic attractor forcing and the merging with chaotic transients under significant levels of forcing. The behaviour of chaotic attractors which combine with chaotic transients from a boundary crisis in the Lorenz-forced model (a two-regime attractor) could be reproduced in the Four-wing model (a four-regime attractor). This is shown in Figure \ref{fig:1-1-2TMergeComp}. We could not find analogous results for either the R\"ossler or Halvorsen examples, both of which are single-regime attractors. Future research in the matter might either validate or disprove such a conjecture with regards to resulting chaotic transients (and ghost attractors in the case of a vanishing basin crisis).

\begin{figure}
    \centering
    \includegraphics[width=0.49\textwidth]{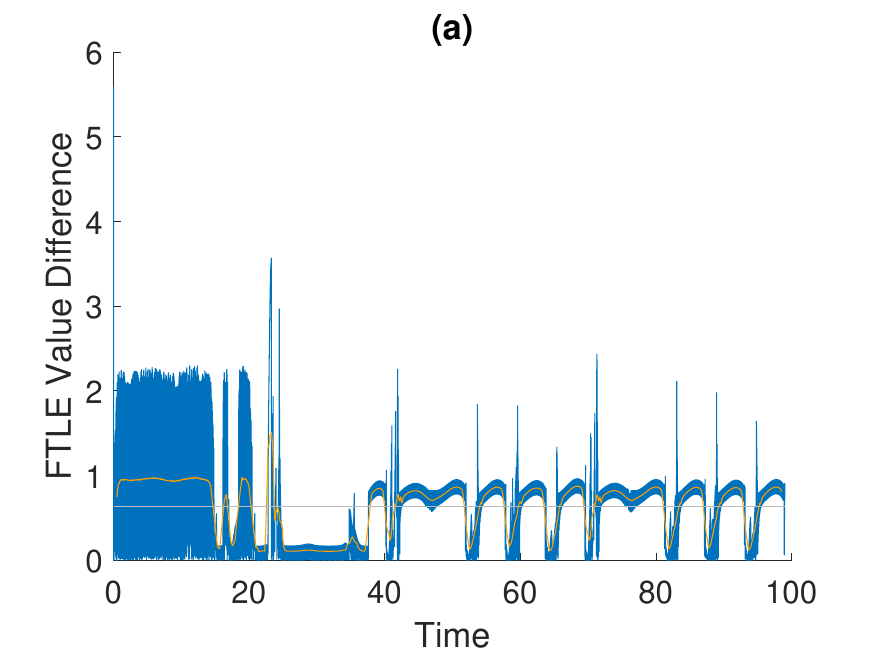}
    \includegraphics[width=0.49\textwidth]{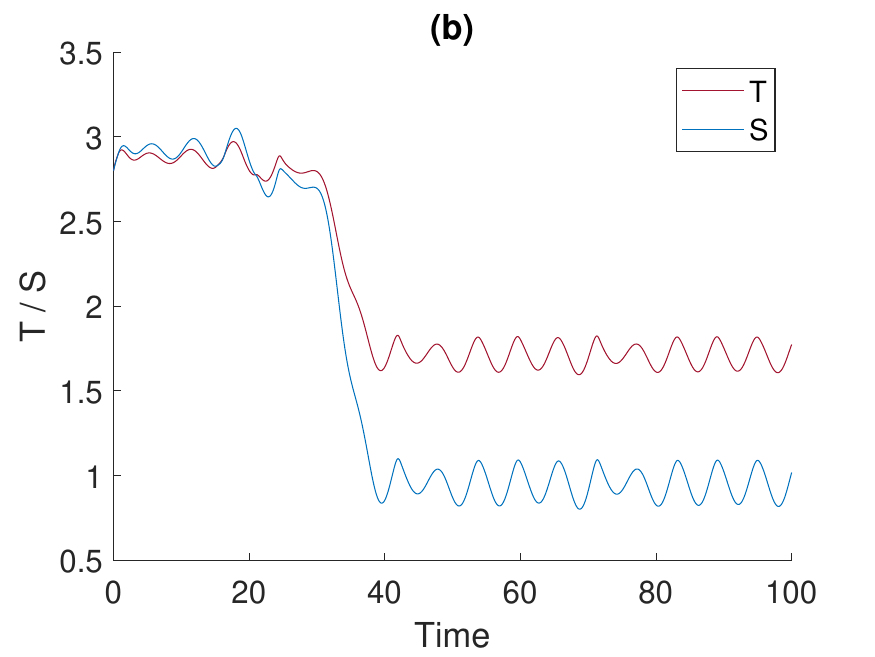}
    \includegraphics[width=0.49\textwidth]{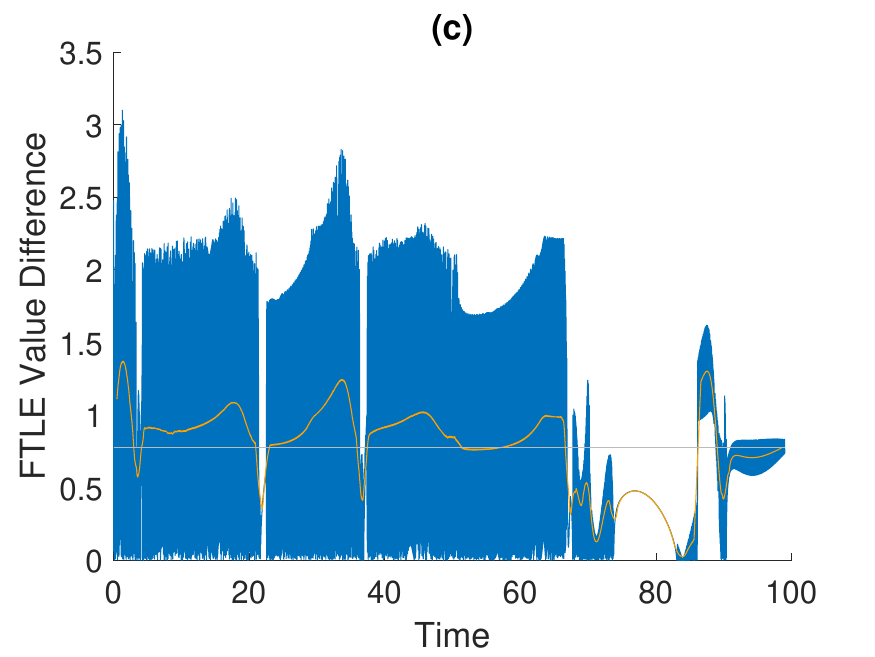}
    \includegraphics[width=0.49\textwidth]{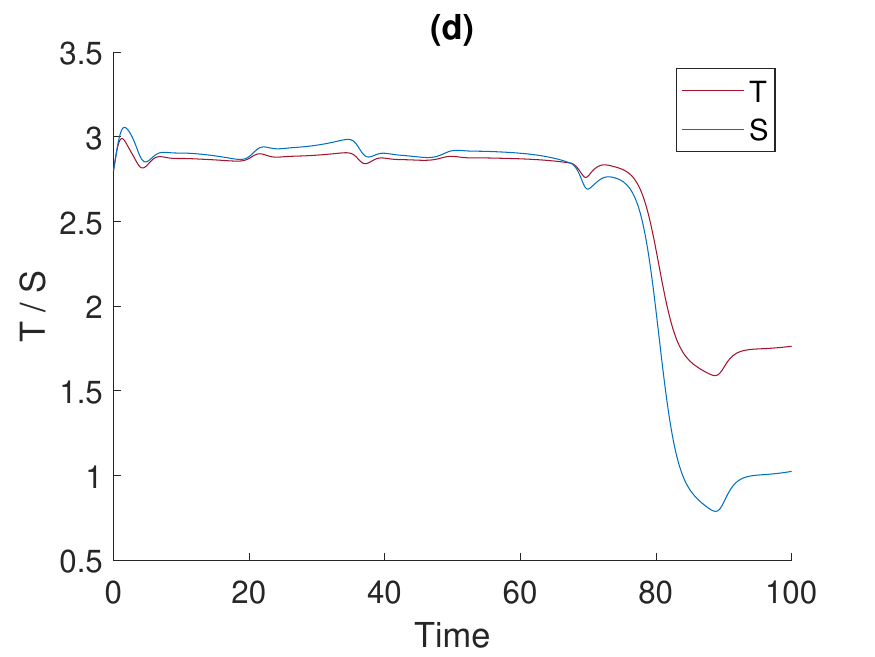}
    \includegraphics[width=0.49\textwidth]{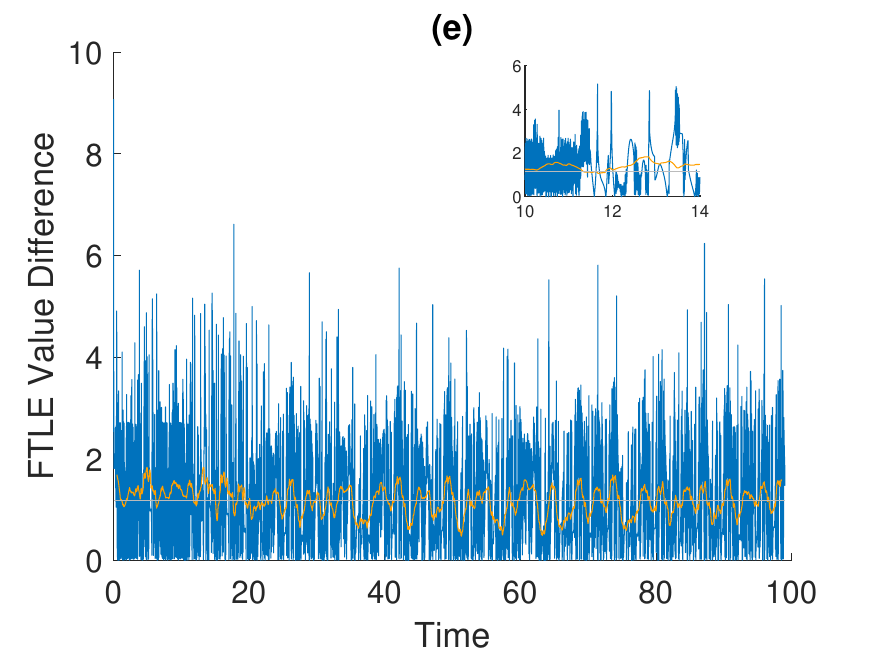}
    \includegraphics[width=0.49\textwidth]{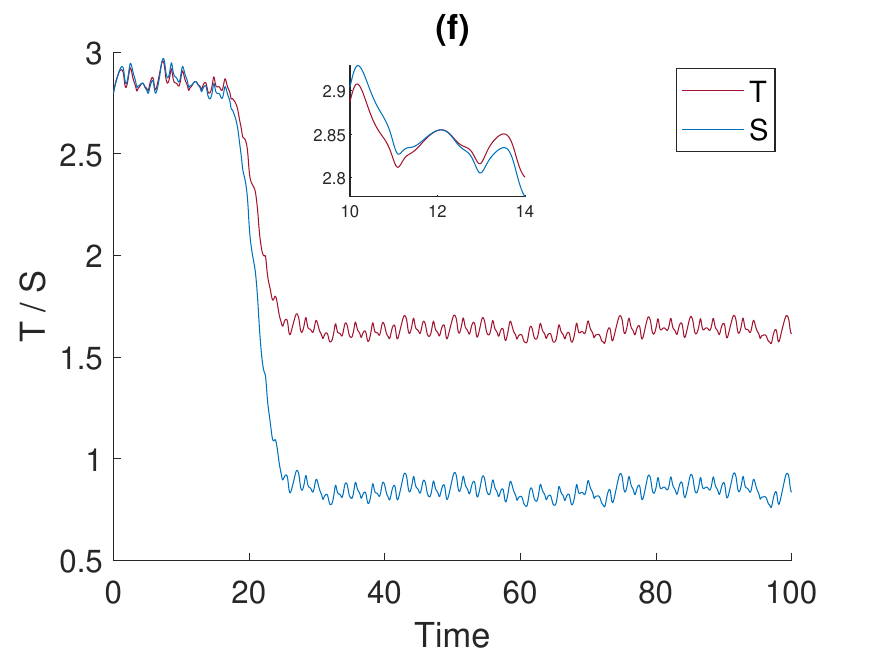}
    \caption{Finite-time analysis of the R\"ossler-forced model (top row, $a$ $=$ $0.0229$), the Four-wing-forced model (middle row, $a$ $=$ $0.0939$) and the Halvorsen-forced model (bottom row  $a$ $=$ $0.0286$). Results of simulations with parameters $(\xi, \eta, \zeta)$ $=$ $(3, 1, 0.3)$ and Stommel initial conditions $(T, S)$ $=$ $(2.8, 2.8)$ are shown. The plots in the left column indicate the distances between the neutral Lorenz and first Stommel FTLEs (with the orange line depicting the rolling average taken over one time unit, and the grey line depicting the average over the full time period). The right column shows the evolution of $T$ and $S$. We find that in all cases we can reproduce the findings of the Lorenz-forced model in terms of decreased distance between the neutral Lorenz and first Stommel FTLE around the SA $\rightarrow$ TH transition.}
    \label{fig:3103GenFTAComp}
\end{figure}

In the Lorenz-forced model the neutral Lorenz and first Stommel FTLEs approach similar values near the point where a trajectory is close to a collision with the chaotic saddle. Finite-time analysis of the other forced models show that this alignment of the first Stommel FTLE and the equivalent neutral FTLE is also present in those models (Figure \ref{fig:3103GenFTAComp}). We observe a drop the absolute distance metric (\ref{eqn:1Norm}) around the point where the crisis occurs. Here, the crisis point is defined by the crossing of trajectories as this implies the attractor has transitioned from SA to TH.

 \section{Conclusion \label{sec:conclusion}}

The Lorenz-Forced Stommel Box Model LFSBM, as defined by (\ref{eqn:LFSBM}), is a recent hybrid model introduced by Ashwin and Newman in their study on measures for pullback attractors and tipping point probabilities \cite{ashwin2021physical}. We have explored solutions to the autonomous forced model in parameter regions that are bistable under no forcing, and have paid particular attention to forcing strengths around levels that induce a crisis in the resulting chaotic attractors. We have found two types of crises in the model: a boundary crisis in which a chaotic attractor is destroyed following a collision with a chaotic saddle, and a vanishing basin crisis that results in the basin of attraction for a chaotic attractor shrinking with increasing forcing. The vanishing basin crisis is a newly discovered type of boundary crisis and results in the existence of a ghost attractor. We also find that further increases to the forcing strength can cause the surviving chaotic attractor to merge with either a chaotic transient (for a typical boundary crisis) or a ghost attractor (for a vanishing basin crisis).

Performing finite-time analysis on the LFSBM reveals that with no forcing the Stommel FTLEs will either oscillate between two specific values or converge monotonically depending on the eigenvalues of the attractor. With forcing present, these Stommel FTLEs start to behave more chaotically. When a crisis is imminent, we find that the first Stommel FTLEs behaves like the neutral Lorenz FTLE around the point where a solution trajectory is at its most sensitive to collision with the chaotic saddle. For the vanishing basin crisis, these alignment of behaviour remains strong even post-crisis.

Our experiments in which we forced the Stommel model with other strange attractors enable us to draw some conclusions as to the possible generality of our findings. We can replicate the approach of the first Stommel FTLE to the neutral strange attractor FTLE around critical transition times. The observations regarding attractors merging with transients and the vanishing of basin crises do depend on the strange attractor used to force the system. We suggest that whether an attractor merges with a chaotic transient (or a ghost attractor) depends on the regime count of the strange attractor used (where two or more regimes in the strange attractor are a requirement). A possible explanation for the vanishing basin crisis that is seen in both the Lorenz-forced and Halvorsen-forced model is currently not apparent.

Possible avenues for future research in the area include the use of Lyapunov vectors to better understand the observed FTLE alignment just before a crisis. It should be verified that this decreasing distance between the neutral and weakly stable FTLEs corresponds to the alignment of the associated Lyapunov vectors. This would imply there is a loss of hyperbolicity when an attractor undergoes a crisis.

\ack
We thank the Sydney Dynamics Group and attendees of the 2022 meeting in Auckland for some insightful discussions.

\appendix

\section{Numerical Methods \label{app:NumMeth}} 
In this Appendix we outline the techniques we employed to generate the results described in the body of the paper. All calculations were performed using MATLAB R2022a and invoked standard ODE solvers to compute the solution trajectories, calculate Lyapunov exponents, and estimate the basin of attraction for a given (chaotic) attractor. We used a fourth-order Runge-Kutta method with the time-step kept constant.

\subsection{Lyapunov Exponents}

The Lyapunov exponents for a given trajectory were calculated using the backwards QR method with Gram-Schmidt orthogonalisation.
We begin with a prescribed interval over which we want to calculate our exponents $t$ $\in$ [$T_0$, $T_m$], a time-step $\Delta t$ and a tolerance value $\varepsilon$.  We computed the solution trajectory over the given time interval using $N$ $=$ $(T_m - T_0)/\Delta t$ steps. We compute the Jacobian $J$ and evaluate the exponential matrix $A$ $=$ $e^{\Delta t J}$ for each time step. The matrix $Q$ is initialized as the identity matrix.

The main calculation was executed using a backwards QR algorithm; see, for example, \cite{dieci1997compuation}. We perform Gram-Schmidt orthogonalisation on $AQ$ after every time step and retain the diagonal elements of $R$ with updated $Q$. After $N$ time steps, we take the average of the logarithm of the diagonal elements of $R$ and divide by the time step to define the Lyapunov spectrum. The iterative process is given such that
\begin{equation}
    Q_{i}R_{i} = A_{i}Q_{i-1},
    \label{eqn:QRDecomp}
\end{equation}
with each $R_i$ stored. Each Lyapunov exponent is calculated via
\begin{equation}
    \lambda_j = \frac{1}{\Delta t N} \sum_{1 = 1}^{N} \ln{R_{i,jj}},
\end{equation}
where $j$ represents the $j^{th}$ row of the spectrum vector. For this study, we used the Gram-Schmidt Orthogonalisation code available online from web.mit.edu \cite{MIT:GSO}.

\subsection{Basins of Attraction \label{app:BoA}}
After computing the Lyapunov exponent over the full time interval [$T_0$, $T_m$] (noting that this converges to the asymptotic exponent as $T_m$ $\rightarrow$ $\infty$), we use it to calculate the basin diagrams; a graphical output of the basin of attraction for a given chaotic attractor. This method, as piloted by Armiyoon and Wu \cite{armiyoon2015novel}, takes advantage of an invariance property that Lyapunov exponents have on a given chaotic attractor to efficiently compute the numerical basin of attraction for some given chaotic attractors. While Armiyoon and Wu \cite{armiyoon2015novel} used Monte Carlo techniques to mitigate the run-time for finding basins of attraction, we instead optimized by limiting our range to the most relevant sections of the Stommel phase plane (either $T_0$, $S_0$ $\in$ [$-3$, $3$], or $T_0$, $S_0$ $\in$ [$0$, $6$]). We then calculated the asymptotic Lyapunov exponents over the sufficiently long finite interval. 

The grid of initial conditions was discretized through the use of a single variable for the step parameter. A smaller parameter gives a higher resolution, but the price to be paid is a commensurate increase in the computational time. Each initial condition is then assigned a color decided by comparing the resulting Lyapunov spectrum to the current list of attractor spectra (intialized as an empty set). If the Lyapunov spectrum is not present in the list of spectra, then it is considered a new chaotic attractor. Otherwise, it is assigned the appropriate color corresponding to the existing spectrum. We expected a maximum of three attractors in any given basin. Owing to the invariance property of Lyapunov exponents, we could minimize the time spent calculating the spectrum for each initial condition by restricting the computation to the latter portion of the trajectory. Once all the initial conditions had been considered, we then plotted the basins of attraction and for each attractor according to assigned color and overlaid the chaotic attractor to the basin diagram.

\subsection{FTLEs}
To calculate the FTLEs along a given solution trajectory we took a windowed approach. We define $I$ to denote the length of each finite-time interval. Given the interval over which we wished to calculate the trajectory $T_m$, the length of a given time step $\Delta t$, and the interval for an individual FTLE calculation $I$, we calculate the FTLEs in the interval $[T_0, T_0 + I]$. We then incremented the window by $\Delta t$ and calculate the FTLEs over the interval $[T_0 + \Delta t, T_0 + \Delta t + I]$. We continue to increment by $\Delta t$ until we reached the final interval of calculation: $[T_m - I, T_m]$. At the conclusion of the process, we returned a matrix of FTLEs in the order as given by the backwards QR method.

\section*{References} 

\bibliographystyle{unsrt}
\bibliography{references}

\begin{thebibliography}{10}

\bibitem{stommel1961thermohaline}
Henry Stommel.
\newblock Thermohaline convection with two stable regimes of flow.
\newblock {\em Tellus}, 13(2):224--230, 1961.

\bibitem{lorenz1963deterministic}
Edward~N Lorenz.
\newblock Deterministic nonperiodic flow.
\newblock {\em Journal of atmospheric sciences}, 20(2):130--141, 1963.

\bibitem{ashwin2021physical}
Peter Ashwin and Julian Newman.
\newblock Physical invariant measures and tipping probabilities for chaotic attractors of asymptotically autonomous systems.
\newblock {\em The European Physical Journal Special Topics}, 230(16):3235--3248, 2021.

\bibitem{dijkstra2005low}
HA~Dijkstra and M~Ghil.
\newblock Low-frequency variability of the large-scale ocean circulation: a dynamical systems approach.
\newblock {\em Reviews of Geophysics}, 43(3):RG3002, 1--38, 2005.

\bibitem{lohmann1999dynamics}
Gerrit Lohmann and Joachim Schneider.
\newblock Dynamics and predictability of {S}tommel's box model. a phase-space perspective with implications for decadal climate variability.
\newblock {\em Tellus A}, 51(2):326--336, 1999.

\bibitem{ben2021useful}
Alona Ben-Tal.
\newblock Useful {Transformations} from {Non}-autonomous to {Autonomous} {Systems}.
\newblock In {\em Physics of Biological Oscillators}, pages 163--174. Springer, 2021.

\bibitem{mehra1996maximal}
Vishal Mehra and Ramakrishna Ramaswamy.
\newblock Maximal {Lyapunov} exponent at crises.
\newblock {\em Physical Review E}, 53(4):3420--3424, 1996.

\bibitem{battelino1988multiple}
Peter~M Battelino, Celso Grebogi, Edward Ott, James~A Yorke, and Ellen~D Yorke.
\newblock Multiple coexisting attractors, basin boundaries and basic sets.
\newblock {\em Physica D: Nonlinear Phenomena}, 32(2):296--305, 1988.

\bibitem{nusse2012dynamics}
Helena~E Nusse and James~A Yorke.
\newblock {\em Dynamics: numerical explorations: accompanying computer program dynamics}, volume 101.
\newblock Springer, 2012.

\bibitem{wagemakers2020saddle}
Alexandre Wagemakers, Alvar Daza, and Miguel~AF Sanju{\'a}n.
\newblock The saddle-straddle method to test for {Wada} basins.
\newblock {\em Communications in Nonlinear Science and Numerical Simulation}, 84:105167, 1--8, 2020.

\bibitem{quinn2019effects}
Courtney Quinn, Jan Sieber, and Anna~S von~der Heydt.
\newblock Effects of periodic forcing on a paleoclimate delay model.
\newblock {\em SIAM Journal on Applied Dynamical Systems}, 18(2):1060--1077, 2019.

\bibitem{quinn2018mid}
Courtney Quinn, Jan Sieber, Anna~S von~der Heydt, and Timothy~M Lenton.
\newblock The {Mid}-{Pleistocene} {Transition} induced by delayed feedback and bistability.
\newblock {\em Dynamics and Statistics of the Climate System}, 3(1):1--17, 2018.

\bibitem{goh2022delayed}
Ryan Goh, Tasso~J Kaper, and Theodore Vo.
\newblock Delayed {Hopf} bifurcation and space--time buffer curves in the complex {Ginzburg}--{Landau} equation.
\newblock {\em IMA Journal of Applied Mathematics}, 87(2):131--186, 2022.

\bibitem{belykh2013multistable}
Igor Belykh, Vladimir Belykh, Russell Jeter, and Martin Hasler.
\newblock Multistable randomly switching oscillators: The odds of meeting a ghost.
\newblock {\em The European Physical Journal Special Topics}, 222(10):2497--2507, 2013.

\bibitem{alkhayuon2019}
Hassan Alkhayuon, Peter Ashwin, Laura~C Jackson, Courtney Quinn, and Richard~A Wood.
\newblock Basin bifurcations, oscillatory instability and rate-induced thresholds for {Atlantic} meridional overturning circulation in a global oceanic box model.
\newblock {\em Proceedings of the Royal Society A}, 475(2225):20190051, 1--26, 2019.

\bibitem{quinn2020}
Courtney Quinn, Terence~J O'Kane, and Vassili Kitsios.
\newblock Application of a local attractor dimension to reduced space strongly coupled data assimilation for chaotic multiscale systems.
\newblock {\em Nonlinear Processes in Geophysics}, 27(1):51--74, 2020.

\bibitem{quinn2021}
Courtney Quinn, Dylan Harries, and Terence~J O’Kane.
\newblock Dynamical analysis of a reduced model for the {North} {Atlantic} {Oscillation}.
\newblock {\em Journal of the Atmospheric Sciences}, 78(5):1647--1671, 2021.

\bibitem{maiocchi2024}
Chiara~Cecilia Maiocchi, Valerio Lucarini, Andrey Gritsun, and Yuzuru Sato.
\newblock Heterogeneity of the attractor of the lorenz’96 model: Lyapunov analysis, unstable periodic orbits, and shadowing properties.
\newblock {\em Physica D: Nonlinear Phenomena}, 457:133970, 1--17, 2024.

\bibitem{ginelli2007characterizing}
Francesco Ginelli, Pietro Poggi, Alessio Turchi, Hugues Chat{\'e}, Roberto Livi, and Antonio Politi.
\newblock Characterizing dynamics with covariant {Lyapunov} vectors.
\newblock {\em Physical Review Letters}, 99(13):130601, 1--4, 2007.

\bibitem{beims2016alignment}
Marcus~W Beims and Jason~AC Gallas.
\newblock Alignment of {Lyapunov} vectors: A quantitative criterion to predict catastrophes?
\newblock {\em Scientific Reports}, 6(1):1--7, 2016.

\bibitem{sharafi2017critical}
Nahal Sharafi, Marc Timme, and Sarah Hallerberg.
\newblock Critical transitions and perturbation growth directions.
\newblock {\em Physical Review E}, 96(3):032220, 1--13, 2017.

\bibitem{letellier2006rossler}
Christophe Letellier and Otto~E Rossler.
\newblock Rossler attractor.
\newblock {\em Scholarpedia}, 1(10), 2006.

\bibitem{wang20093}
Zenghui Wang, Yanxia Sun, Barend~Jacobus van Wyk, Guoyuan Qi, and Michael~Antonie van Wyk.
\newblock A 3-d four-wing attractor and its analysis.
\newblock {\em Brazilian Journal of Physics}, 39:547--553, 2009.

\bibitem{vaidyanathan2016adaptive}
Sundarapandian Vaidyanathan and Ahmad~Taher Azar.
\newblock Adaptive control and synchronization of {Halvorsen} circulant chaotic systems.
\newblock In {\em Advances in chaos theory and intelligent control}, pages 225--247. Springer, 2016.

\bibitem{dieci1997compuation}
Luca Dieci, Robert~D Russell, and Erik~S Van~Vleck.
\newblock On the computation of {Lyapunov} exponents for continuous dynamical systems.
\newblock {\em SIAM Journal on Numerical Analysis}, 34(1):402--423, 1997.

\bibitem{MIT:GSO}
Massachusetts~Institute of~Technology~Essays.
\newblock Gram-{Schmidt} in 9 lines of {M}{A}{T}{L}{A}{B}.
\newblock {\em http://web.mit.edu/18.06/www/Essays/gramschmidtmat.pdf}, 2022.

\bibitem{armiyoon2015novel}
Ali~Reza Armiyoon and Christine~Q Wu.
\newblock A novel method to identify boundaries of basins of attraction in a dynamical system using {Lyapunov} exponents and {Monte} {Carlo} techniques.
\newblock {\em Nonlinear Dynamics}, 79(1):275--293, 2015.

\end{thebibliography}

\end{document}